\documentclass[traditabstract]{aa} % for the abstract without structuration 
\usepackage{graphicx}
\usepackage{txfonts}
\usepackage{natbib}
\usepackage{lscape}
\bibpunct{(}{)}{;}{a}{}{,} % to follow the A&A style
\def\ltsima{$\; \buildrel < \over \sim \;$}
\def\simlt{\lower.5ex\hbox{\ltsima}}
\def\gtsima{$\; \buildrel > \over \sim \;$}
\def\simgt{\lower.5ex\hbox{\gtsima}}
\def\SN{\hbox{S/N}}
\def\teff{\mathrm{T_\mathrm{eff}}}

\def\kms{\mbox{km~s$^{-1}$}}

\begin{document}
%\titlerunning{Distance determination for RAVE stars}
\authorrunning{Zwitter et al.}


\title{Distance determination for RAVE stars using stellar models. II}
\subtitle{Most likely values assuming a standard stellar evolution scenario}

\author{
T. Zwitter\inst{1,2},
G. Matijevi\v{c}\inst{1},
M. A. Breddels\inst{3},
M. C. Smith\inst{3,4},
A. Helmi\inst{3},
U. Munari\inst{5},
O. Bienaym\'{e}\inst{6},
J. Binney\inst{7},
J. Bland-Hawthorn\inst{8},
C. Boeche\inst{9},
A. G. A. Brown\inst{10},
R. Campbell\inst{11},
K. C. Freeman\inst{12},
J. Fulbright\inst{13},
B. Gibson\inst{14},
G. Gilmore\inst{15},
E. K. Grebel\inst{16},
J. F. Navarro\inst{17},
Q. A. Parker\inst{18},
G. M. Seabroke\inst{19},
A. Siebert\inst{6},
A. Siviero\inst{5,9},
M. Steinmetz\inst{9},
F. G. Watson\inst{20},
M. Williams\inst{9},
R. F. G. Wyse\inst{13}
}
%
%% here is the list of institutes
\institute{
{University of Ljubljana, Faculty of Mathematics and Physics, Ljubljana, Slovenia}
\and {Center of excellence SPACE-SI, Ljubljana, Slovenia}
\and {Kapteyn Astronomical Institute, University of Groningen, Groningen, The Netherlands}
\and {Kavli Institute for Astronomy and Astrophysics, Peking University, Beijing, China}
\and {INAF Astronomical Observatory of Padova, 36012 Asiago (VI), Italy}
\and {Observatoire de Strasbourg, Strasbourg, France}
\and {Rudolf Peierls Centre for Theoretical Physics, University of Oxford, UK}
\and {Sydney Institute for Astronomy, School of Physics, 
University of Sydney, NSW 2006, Australia}
\and {Astrophysikalisches Institut Potsdam, Potsdam, Germany}
\and {Leiden Observatory, Leiden University, Leiden, The Netherlands}
\and {Western Kentucky University, Bowling Green, Kentucky, USA}
\and {RSAA, Australian National University, Canberra, Australia}
\and {John Hopkins University, Baltimore, Maryland, USA}
\and {University of Central Lancashire, Preston, UK}
\and {Institute of Astronomy, Cambridge, UK}
\and {ARI, Zentrum f\"ur Astronomie der Universit\"at Heidelberg, Heidelberg, Germany}
\and {University of Victoria, Victoria, Canada}
\and {Macquarie University, Sydney, Australia}
\and {e2v Centre for Electronic Imaging,
Planetary and Space Sciences Research Institute, The Open University, Milton
Keynes, UK}
\and {Anglo Australian Observatory, Sydney, Australia}
}



\date{Received 3 May 2010 / Accepted 26 June 2010}

\abstract{
The RAdial Velocity Experiment (RAVE) is a spectroscopic survey of the 
Milky Way which already collected over $400\,000$ spectra of $\sim 330\,000$ 
different stars. We use the subsample of spectra with spectroscopically 
determined values of stellar parameters to determine 
the distances to these stars. The list currently contains 
$235\,064$ high quality spectra which show no peculiarities and belong to 
$210\,872$ different stars. The numbers will grow as the RAVE 
survey progresses. The public version of the catalog will be made 
available through the CDS services along with the ongoing RAVE 
public data releases. 

The distances are determined with a method 
based on the work by Breddels et al.~(2010). Here we assume that the star 
undergoes a standard stellar evolution and that its spectrum 
shows no peculiarities. The refinements include:  
the use of either of the three isochrone sets, a better account 
of the stellar ages and masses, use of more realistic errors of
stellar parameter values, and application to a larger dataset. 
The derived distances of both dwarfs and giants match within 
$\sim 21$\% to the astrometric distances of Hipparcos 
stars and to the distances of observed members of 
open and globular clusters. Multiple observations of a fraction of 
RAVE stars show that repeatability of the derived distances is even 
better, with half of the objects showing a distance scatter 
of  $\simlt 11$\%. 

RAVE dwarfs are $\sim 300$~pc from the Sun, and giants are at distances 
of 1 to 2~kpc, and up to 10~kpc. This places the RAVE dataset between 
the more local Geneva-Copenhagen survey and the more distant and 
fainter SDSS sample. As such it is ideal to address some of the 
fundamental questions of Galactic structure and evolution in the 
pre-Gaia era. Individual applications are left to separate papers, 
here we show that the full 6-dimensional information on position and 
velocity is accurate enough to discuss the vertical structure 
and kinematic properties of the thin and thick disks. 
}


\maketitle
%%%%%%%%%%%%%%%%%%%%%%%%%%%%%%%%%%%%%%%%%%%%%%%%%%%%%%%%%%%%%%%%%%%%%%%%%%%%
%%%%%%%%%%%%%%%%%%%%%%%%%%%%%%%%%%%%%%%%%%%%%%%%%%%%%%%%%%%%%%%%%%%%%%%%%%%%

\section{Introduction}
\label{s:introduction}

It is now widely accepted that our Galaxy can be used as a testbed to study
galaxy formation and evolution \citep{FBH02}.
A vital tool for this goal is the ability 
to use complete information on position, motion and physical properties of a 
large number of individual stars. Some of the quantities are usually known, 
i.e.\ the sky coordinates and in our case proper motions. But characterizing 
the three-dimensional position requires knowledge of the distance to the 
star and describing its motion requires knowledge of its radial velocity.
So one needs to measure also these quantities in order to have complete 
information of an object's place in the 6-dimensional phase space
which combines the position and velocity vectors. 
RAVE (RAdial Velocity Experiment) is the only large scale survey 
measuring radial velocities spectroscopically, with the exception of 
the SDSS survey which mainly samples the halo of the Galaxy. 
RAVE already collected over $400\,000$ spectra of $\sim 330\,000$ different 
stars. Most of the RAVE objects are thin and thick disk stars. So the SDSS 
and RAVE surveys are complementary in the sample surveyed (halo vs.\ disk), 
magnitude range (fainter vs. brighter) and even the celestial hemisphere 
(North vs. South). The RAVE survey is described in the articles accompanying 
the first two public data releases \citep[][ hereafter Z08]{S06,Z08}. 
The article describing the third public release is in preparation
\citep{siebert10}.  
With the exception of the first data release the subsequent ones also 
include results of spectroscopic measurements of stellar physical 
parameters: the effective temperature, surface gravity, and metallicity.

The RAVE dataset can be used to study a range of scientific topics 
\citep{steinmetz2003}, including present structure and formation 
of the Galaxy, remnants of recent accretion events, and discovery of 
individual peculiar objects and spectroscopic binary stars \citep{matijevic10}. 
It was already used to better constrain the Galactic escape speed  
at the Solar radius 
\citep{escape_speed}. \citet{veltz} studied kinematics towards the 
Galactic poles and identified discontinuities that separate thin disk, 
thick disk and a hotter component. \citet{seabroke2008} searched for 
in-falling stellar streams on to the local Milky Way, and 
\citet{siebert08} estimated the tilt of the stellar velocity ellipsoid 
and its implications for mass models.
\citet{pasetto10} studied the kinematics of the thick disk.
 \citet{munari_DIBS}
showed that a diffuse interstellar band at 8620~\AA\ can be used to 
reliably measure the amount of galactic interstellar reddening. 
\citet{munari_LBV} studied the properties of Luminous Blue Variables 
in the Large Magellanic Cloud. They used a photo-ionization analysis 
of their rich emission line spectra to demonstrate the great 
diagnostic potential of RAVE in spectroscopy of peculiar stars.
Additional ongoing studies have been listed 
in \citet{S06}. 
 
Most of these studies would profit from the knowledge of stellar distance.
With very few exceptions, RAVE stars are too far away and too faint to have 
trigonometric parallaxes measured by the Hipparcos mission.
The RAVE sample is magnitude limited
($9 < I_{\rm DENIS} < 13$) but otherwise entirely randomly selected from the Tycho 
(bright stars), SuperCOSMOS (faint stars) and DENIS catalogs. So only a 
small fraction of objects observed by RAVE are members of stellar clusters 
or are of special type for which standard candle techniques can be used 
to judge their distances. One choice is to assume that all RAVE stars
belong to the main sequence and to use their photometric colors to 
infer their distance \citep{klement08}. Unfortunately, giants form 
about half of the RAVE sample, so any conclusions based on the assumption 
that they are main sequence stars will be biased. Another possibility is 
to use photometric colors to isolate the red clump stars. This is a 
viable technique, but it applies only to a fraction of the sample and 
may be influenced by contaminating outliers of similar colors. 
 
Over 90\% of the RAVE spectra belong to normal stars and have no substantial 
instrumental problems. So one may use the knowledge of spectroscopically 
measured physical parameters to infer their abolute position on the 
H-R diagram and thus their absolute magnitude. The choice of a near infrared 
passband ($J_{\rm2MASS}$) and the fact that the vast majority of RAVE stars 
lie at high Galactic latitudes ($|b|>20^\mathrm{o}$) means that the influence
of interstellar reddening is negligible. This approach was introduced 
by our first paper on the subject 
\citep[][ hereafter B10]{B10}. Here we build on this solid foundation and 
refine this approach by a better accounting for measurement errors, 
stellar evolution knowledge, allowing for three different sets of 
stellar isochrone computations, a refined computation method, and 
by expanding the sample to the recently observed stars. Another distance 
computation method using a Bayesian approach tested on a Galactic 
pseudo-data sample will also be used to calculate distances to RAVE stars  
\citep{burnett10}.

All these efforts are aimed at a common goal to derive the most reliable 
distances to RAVE stars. The goal is important and 
it is likely that different distance determination methods will be 
preferred for different types of stars. The situation could indeed prove  
to resemble the problem of computation of stellar isochrones where it is 
best for the user to choose a particular set depending on the problem 
which is being considered.

By the end of the decade a public catalog of extremely accurate and 
numerous astrometric distances measured by the ESA's mission 
Gaia is expected to become available. Still, as with any magnitude limited 
survey, there will be a significant portion of stars for which their 
distances are too large for accurate trigonometric parallaxes to be 
determined. So photometric distance determination methods will remain to 
be useful also in the post-Gaia era of Galactic astronomy.

The structure of the paper is as follows: Sec.~\ref{s:method} introduces the 
refined distance determination method. In Sec.~\ref{s:tests} we test the 
results by comparing them to the available Hipparcos trigonometric 
parallax measurements and to the distances of open clusters with 
some confirmed members observed by RAVE. Sec.~\ref{s:results}
discusses statistical properties of the results and the structure of 
the publicly available catalog, to be followed by general 
conclusions in Sec.~\ref{s:conclusions}. The reader is encouraged to 
consult B10 for a more complete discussion of different 
methods to measure stellar distances and some of the scientific results. 
Detailed scientific results related to the structure and evolution 
of the Galaxy as well as distances of individual stars and stellar clusters 
will be published separately.

\section{Refined distance determination method}
\label{s:method}

B10 used values of stellar parameters and the 
Yonsei-Yale \citep{demarque2004} isochrones to derive the absolute 
$J_\mathrm{2MASS}$ magnitudes. Comparison with the apparent 2MASS $J$ 
magnitudes then yielded distances. The values of stellar parameters 
measured by RAVE are not error-free. B10 assumed constant errors in 
temperature (300~K), gravity (0.3~dex), and metallicity (0.25~dex) 
for all stars. So  the values derived by RAVE were jittered according 
to these assumed standard uncertainties.  5000 realisations of each 
observation were calculated by sampling Gaussian 
distributions in each observable. The final absolute $J$ magnitude
was the mean of the absolute J magnitudes pertaining to the closest 
isochrone matches for each of the realisations. The closest match 
was derived by minimizing the usual $\chi^2$ statistics. The distribution 
of absolute $J$ magnitudes for individual realisations also gave 
the standard error of the derived absolute magnitude and thus of the 
distance to the star. 

In formulating refinements to the B10 approach we alter the 
method for distance derivation. Instead of using the closest
match, an expectation value of the absolute magnitude is derived 
considering several isochrones at a given metallicity. Our goal is 
to derive the {\it most likely} distance to the star. So we assume  
that we are dealing with a normal star that follows stellar 
evolution as mirrored by theoretical isochrones. We use 
our knowledge of stellar evolution to compensate for some of the 
errors in values of stellar parameters as derived by RAVE. So the 
results of this paper should be very useful in Galactic and general 
population studies. But we note that our results may not be applicable
to peculiar objects or to those in rapid transit phases of stellar 
evolution. The latter because the adjacent longer lived and so 
more populated evolutionary phases may be within the measurement 
errors. The refinements to the B10 method are described below.

\subsection{The three sets of isochrones}

The distance computation of B10 was based on the Yonsei-Yale   
\citep[hereafter YY,][]{demarque2004} set of isochrones. Here we add two additional 
sets: the Dartmouth \citep{dotter08} and the Padova isochrones 
\citep{bertelli08}. While the former two sets are very useful 
for the derivation of properties of stars which are on or close to the 
main sequence, their computations do not extend to the asymptotic giant branch 
(AGB) or red clump (RC) evolutionary phases. The Padova set on the other 
hand nicely covers both the AGB and the RC phase.
So it is particularly useful to derive 
distances to giant stars. All sets of isochrones tabulate also the 
absolute $J$ magnitude. In the case of YY and Dartmouth sets 
the absolute $J$ magnitude can be directly compared to the apparent 
one measured by the 2MASS survey, while eq.~1 from \citet{koen07}
was used to transform the Padova set into the 2MASS system. In all 
cases we used isochrones without any enhancement of alpha elements. 
This quantity is not measurable reliably by RAVE and has a very small  
influence on the final distance results (see B10 for a detailed discussion). 

All isochrone sets cover the parameter space in the necessary detail and 
in some cases even offer interpolation routines to derive values at 
points in between. These routines were used when possible, otherwise 
a standard linear interpolation scheme was used. 
So we generated a fine grid of isochrones with age, mass and metallicity 
steps much smaller than the observational errors. The mass step 
was always smaller than 0.01~M$_\odot$, and the metallicity step was 
0.1~dex. Sampling in age is discussed below.

All computations were performed three times, deriving a separate set of 
distances for each set of isochrones. We publish all three distances 
and their errors. It is then up to the user to decide which one is the 
most suitable for the type of problem or the type of object that is 
being considered. Details of isochrone computation and the corresponding
assumptions are described in detail in the original papers presenting each 
isochrone set.

\subsection{Isochrones uniformly spaced in time}
\label{s:uniform_spacing}

B10 used a set of isochrones with 40 different ages, spaced 
logarithmically between 0.01 and 15.0 Gyr. 
%So the age sequence was as follows: 0.01~Gyr, 0.012~Gyr, 0.015~Gyr, ...
%4.04~Gyr, 4.87~Gyr, 5,87~Gyr, ..., 10.31~Gyr, 12.44~Gyr, 15~Gyr.
This implies a time step of only 2~Myr at the young end and up to
2.5~Gyr at the old end of the isochrone set. Such a choice 
covers the [temperature, gravity] plane well. But the use of the 
closest match implies an implausibly high probability that a 
certain realisation lies within a rapid evolutionary phase, such 
as a contraction towards the main sequence or a transit from the 
main sequence towards the region of giants. 
The bulk of the Galactic populations is expected to be found in 
slow, long--lasting phases of stellar evolution. So, although it may 
be possible that RAVE observed some objects in rapid phases of 
stellar evolution, this is not probable in individual cases. 

In the present calculations the isochrones are spaced uniformly and not 
logarithmically in time. So the density of grid points is inversely 
proportional to the speed of motion of a star across the 
[temperature, gravity] plane. We use a step of 20 Myr 
for ages up to 12~Gyr. The youngest age depends on its availability 
for a particular isochrone set. It is 0.1~Gyr for the YY set, 
1~Gyr for Dartmouth and 0.5~Gyr for the Padova one. Uniform spacing in
time gives a proper weight to longer 
evolutionary phases, while rapid transit periods have their 
importance diminished. The adopted age range includes the span 
of ages of a general stellar population as determined by the 
Geneva-Copenhagen survey \citep{holmberg09}.

\subsection{Weighting according to mass and absolute magnitude}
\label{s:x}

The quantity changing along a theoretical isochrone is stellar mass. 
Isochrones are tabulated so that connecting adjacent points follows 
their shape well. So the points are tabulated in relatively large 
mass steps while the isochrone lies on the main sequence. The mass 
steps get much smaller when the isochrone bends toward giants. 

We use a scheme that weights individual solutions. Clearly the weight 
of a given point should be proportional to a mass step $dm$ which is an 
average of mass differences to the adjacent mass points on a given 
isochrone. It should also be noted that high mass stars are much less  
frequent than ones with sub-solar masses. $dN/dm$ is the Galactic 
mass function as given in eq.~17 of \citet{chabrier03} for sub-solar 
stars and in his eq.~11 for stars more massive than the Sun. Moreover,  
more luminous stars at a given apparent magnitude are farther 
away than their less luminous counterparts. The volume of space 
within which the stars with a given absolute magnitude $M_I$ in the 
Cousin $I$ band appear brighter than the given apparent limiting 
magnitude is proportional to $10^{-0.6 M_I}$. We use the $I$ band, 
as both the surveys used to select the observing sample and the final 
RAVE spectra are centered on this band. If we assume that the space 
density of stars does not depend on distance  the following relative 
weight $x$ of points along the isochrone is obtained:
\begin{equation}
x = dm (dN/dm) 10^{-0.6 M_I}
\label{eq:x}
\end{equation}  

One may argue that the $dN/dm$ term can be used only for a volume limited
survey, while RAVE is a magnitude limited one. Similarly, the stellar 
distribution is not uniform. Stars at large distances are quite far away 
from the Galactic plane, so the frequency of these luminous, massive stars 
is even lower than would be in the case of a uniform space density. 
Fortunately these assumptions are not crucial. It is reasonable to 
take into account that massive stars are uncommon and that RAVE is 
able to sample a larger volume of more luminous stars. But the exact 
mathematical form of these prescriptions is not so important, 
and we can choose the simplest relation using varying weights. The reason 
lies in the quality of RAVE's measurement of values of stellar parameters. 
Their errors are rather small, even though not negligible. So any 
large offsets in temperature, gravity and metallicity are virtually 
excluded. The role of equation~\ref{eq:x} is to ascribe a relative weight 
to points on the isochrones that are close to the initial RAVE values. 

\begin{figure}%[hbtp]
\centering
\includegraphics[width=7.1cm,angle=270]{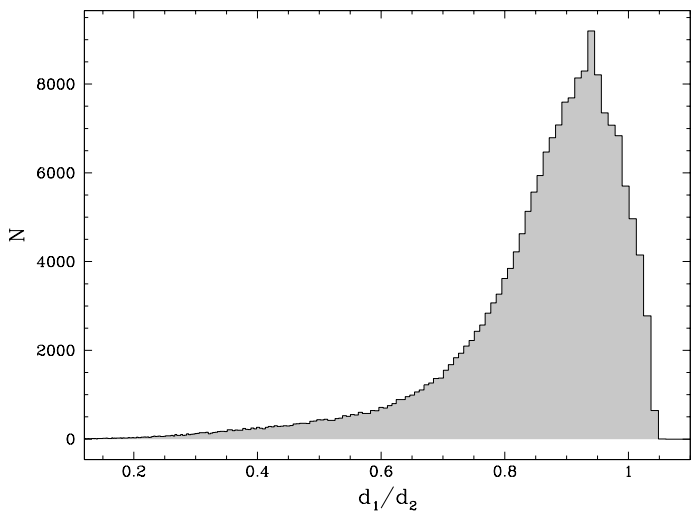}
\caption{
Ratio of distances derived under different assumptions. $d_1$ is 
derived assuming that the probability of a star to be observed does not 
depend on its mass or brightness, while the distance $d_2$ 
takes into account the mass and brightness functions, as described in the text.  
%mode: 0.94 (most probable)
%median: 0.87 (half up half down)
}
\label{FIGplotssm_IMFdiff}
\end{figure}

As a test we recalculated the whole database using $x = dm$. So we 
ignored any mass or luminosity dependence. The results are illustrated 
in Fig.~\ref{FIGplotssm_IMFdiff}. We see that most stars are closer 
than in the case when we use eq.~\ref{eq:x}. The magnitude 
term that favours more luminous and hence more distant stars in 
eq.~\ref{FIGplotssm_IMFdiff} therefore dominates over the mass function 
$dN/dm$ which disfavours massive (and so luminous) stars. But differences 
in distances derived by the test procedure and the usual one are small: 
the mode of the distance ratio distribution derived by the two methods 
is 0.94, and the median is 0.87. So even the extreme simplification 
which ignores any assumptions used to derive eq.~\ref{eq:x}  
changes the distances by only $\sim 10$\%.

\subsection{Assumed errors of stellar parameters}
\label{s:assumed_errors}

The errors of stellar parameter values measured by RAVE depend on stellar
type. Typical temperature errors for stars cooler than 9000~K are around
400~K. The errors are smallest for supergiants. Their
atmospheres are the most transparent ones, so the wealth
of spectral lines arising at different optical depths can improve
the temperature accuracy. Understandably the errors for metal-poor 
stars are larger than for their Solar counterparts. The errors
get considerably worse for stars hotter than 9000~K, where most
metal lines are missing and the spectrum is largely dominated
by hydrogen lines. Similarly, strong wings of hydrogen lines, which
are sensitive to gravity, allow small gravity errors in hot stars.
On the other hand rather narrow metallic lines in the RAVE
wavelength range, including those of Ca~II, do not allow an
accurate determination of gravity in cool stars. The gravity error
in cool stars has a strong gravity dependence: in dwarfs it is 
$\sim 0.6$~dex, but the rather transparent atmospheres of giant 
stars still allow for a more accurate gravity determination, with 
a scatter that can be as low as 0.2~dex. Finally, the typical error 
in metallicity for stars cooler than 7000~K is 0.2~dex. The error 
for hotter stars is understandably much larger, as these stars 
lack most of the metallic lines in their spectra. 

Given this vivid behaviour it seems important to allow for a variation 
of the accuracy of stellar parameter values across the parameter space.
This is a refinement over other studies aiming at distance computation 
using values of stellar parameters as determined by RAVE. Here 
the errors are obtained from a linear interpolation of results presented 
in Fig.~19 and equations 22 and 23 of Z08. Note that these 
error estimates may be rather conservative \citep{S09}.

\subsection{Calculating the distance modulus}

Values of stellar parameters, as determined by RAVE, are not error-free.
So B10 first determine the closest match to the RAVE values in the isochrone 
parameter space. This closest match is then jittered according to the 
assumed parameter uncertainties. Each of the realizations again yields 
the closest match in the isochrone space, and a suitable average of 
these closest matches yields the final value of the distance and its 
scatter. 

This procedure can benefit from a refinement regarding the use of the 
closest match. Imagine we have a Solar type main-sequence star that 
would have the reported values of RAVE parameters placing it just above 
the main sequence of the H-R diagram. Jittering the reported value of 
its gravity would mean that half of the realizations would imply a 
higher gravity and so a main-sequence solution. But the other half 
of the jittered realizations would claim the star is above the main 
sequence in a rapid transition towards the giant phase.  
The use of the closest matches would thus produce half of the 
realizations involving massive stars in a rapid transit toward giants 
and half of main-sequence cases. The relative weight of realizations 
involving the rapid transition phase is therefore too large. It also 
influences the value of the mean of absolute $J$ magnitudes for 
individual realizations and so the derived distance to the star. 

We use a modified procedure to calculate the most likely value of 
the absolute $J$ magnitude.  The distance modulus then follows from 
its difference with the apparent $J$ magnitude as measured by the 
2MASS survey. We do not jitter the measured parameter values and calculate  
individual realizations to be solved with the closest match.
Instead, the expectation value of the absolute magnitude is derived 
using the relative weights 
\begin{equation}
X(a,m) = x(a,m) \,\,\,\, e^{-y(a,m)}\label{eq:X}
\end{equation}
\begin{eqnarray} 
y(a,m) = \frac{[T\mathrm{(RAVE)}-T(a,m)]^2}{2 \sigma^2_{T\mathrm{(RAVE)}}}
+\frac{[\log g \mathrm{(RAVE)}-\log g(a,m)]^2}{2 \sigma^2_{\log g 
\mathrm{(RAVE)}}} \nonumber  
\end{eqnarray}
where $a$ is age and $m$ mass at a given point on the isochrone 
with the metallicity value of the RAVE measurement. Standard 
deviations of the measured value of temperature 
($\sigma_{T\mathrm{(RAVE)}}$) and gravity 
($\sigma_{\log g\mathrm{(RAVE)}}$) are determined 
as described in Sec.~\ref{s:assumed_errors}. The final value 
of the absolute $J$ magnitude ($M_J$) and its standard deviation 
($\sigma(M_J)$) are calculated from the first moments of the distribution
\begin{equation}
M_J = \frac{\sum_a \sum_m M_J(a,m) X(a,m)}{\sum_a \sum_m X(a,m)}
\end{equation}
\begin{equation}
\sigma^2(M_J) = \frac{\sum_a \sum_m X(a,m)[M_J-M_J(a,m)]^2}{\sum_a \sum_m X(a,m)}
\label{eq:sigma}
\end{equation}
The procedure makes a brute force average of all solutions for an 
isochrone set of a metallicity measured by RAVE. Typical values 
of metallicity errors are $\simeq 0.2$~dex, or about half of the 
step in a grid of stellar atmosphere models used by Z08 (their Fig.\ 19).
On the other hand the errors in temperature or gravity are 
typically larger than the grid step. The isochrone 
sets for different values of metallicity are very similar, as 
already discussed by B10 (see its Fig.~2). A lower metallicity 
corresponds to a more transparent stellar atmosphere with a hotter effective 
temperature, but has little effect on the stellar energy output. 
So a metallicity change causes a shift of the isochrone in the temperature 
direction, but the absolute magnitude stays very similar (see 
isochrone comparison in Fig.~\ref{FIGHRHipparcos}b below). 
It is thus not surprizing that the results on absolute magnitude show 
no significant changes when using metallicity as another independent 
variable. In particular, the distances do not change by more than 10\%\ 
for 90\%\ of the spectra when using metallicity as an independent 
variable. But an extra term significantly prolongs the calculation.
%done for 33000 spectra of the Padova set (see the specific directory).
So we decide to adopt the simplified form given in 
Eq.~\ref{eq:X}. If isochrones with the value of 
RAVE-measured metallicity do not exist they are obtained by a linear 
interpolation from the adjacent ones. Similarly, some of 
the isochrone sets are calculated for a set of iron abundances 
([Fe/H]) which in general differ from the metallicities ([M/H]) measured 
by RAVE. Equation 21 from Z08 is used to convert between 
iron abundance and metallicity. Similarly we use eq.~20 from Z08 
to calibrate the metallicity values.
 
We explained 
in Sec.~\ref{s:uniform_spacing} that the isochrones are uniformly 
spaced in age, while Sec.~\ref{s:x} discusses relative weights $x(a,m)$
along an isochrone. A Gaussian form of eq.~\ref{eq:X} means that 
only points on the isochrones that have the temperature and gravity 
within a few standard deviations from the RAVE measurement are important.
This scheme weights each solution according to its distance from the 
position measured by RAVE. So the weighting goes by distance in the 
[temperature, gravity] parameter space and not by area, as is the case 
with the closest match. 
So we avoid giving excessive weights to isochrones in the scarcely 
populated regions of the parameter space, where a given solution 
is the closest match to a substantial range of the [temperature, gravity]
pairs.

\section{Tests of the derived distances}
\label{s:tests}

Stellar distances are generally not known, so we can test our 
results only for certain subsets of the measured sample. Two such 
cases will be considered: Hipparcos stars and confirmed members of 
stellar clusters. 
In future we plan to test the results also with a Galactic pseudo-data 
set \citep{burnett10}. Combined results of these tests and experience 
obtained through a scientific use of distance results will be discussed 
in future releases of RAVE data.

\subsection{Hipparcos stars}
\label{s:Hipparcos}

%\begin{landscape}
\begin{table*}
%\rotate
%\tabletypesize{\small}
\centering
\caption{Comparison of spectroscopic and trigonometric distances $^{(1)}$
}
\label{t:hipparcos}
%\tablewidth{0pt}
%\tablecolumns{7}
%\tablehead{
\begin{tabular}{lcccccc}
\hline\hline
            &\multicolumn{3}{c}{Sample A$^{(2)}$}&\multicolumn{3}{c}{Sample B$^{(3)}$}\\ 
Calculation &$k$ & $\sigma$ &spectra  &$k$ & $\sigma$ &spectra   \\
%}
\hline
%\startdata
B10                              &$0.92 \pm 0.01$ & 0.18 & 277 &$0.89 \pm 0.01$ & 0.18 & 207 \\
This paper (YY isochrones)       &$1.03 \pm 0.01$ & 0.23 & 725 &$1.09 \pm 0.02$ & 0.25 & 207 \\
This paper (Dartmouth isochrones)&$1.01 \pm 0.01$ & 0.23 & 776 &$1.07 \pm 0.02$ & 0.24 & 207 \\
This paper (Padova isochrones)   &$1.03 \pm 0.01$ & 0.25 & 674 &$1.09 \pm 0.02$ & 0.26 & 207 \\
\hline
%\enddata
\end{tabular}
%\tablecomments{
\tablefoot{
(1) Spectroscopic ($d$) and trigonometric ($d_\varpi$) 
distances are assummed to follow the relation $d = k d_\varpi$.
(2) Hipparcos stars with a trigonometric distance error smaller 
than 20\%, a spectroscopic distance error below 32\%\ and with RAVE 
spectra with S$/$N~$>40$. (3) A subset of Sample A where all four 
calculations find a spectroscopic distance error smaller than 32\%.
}
%\end{deluxetable}
\end{table*}
%\end{landscape}

\begin{figure}[hbtp]
\centering
\includegraphics[width=6.9cm,angle=270]{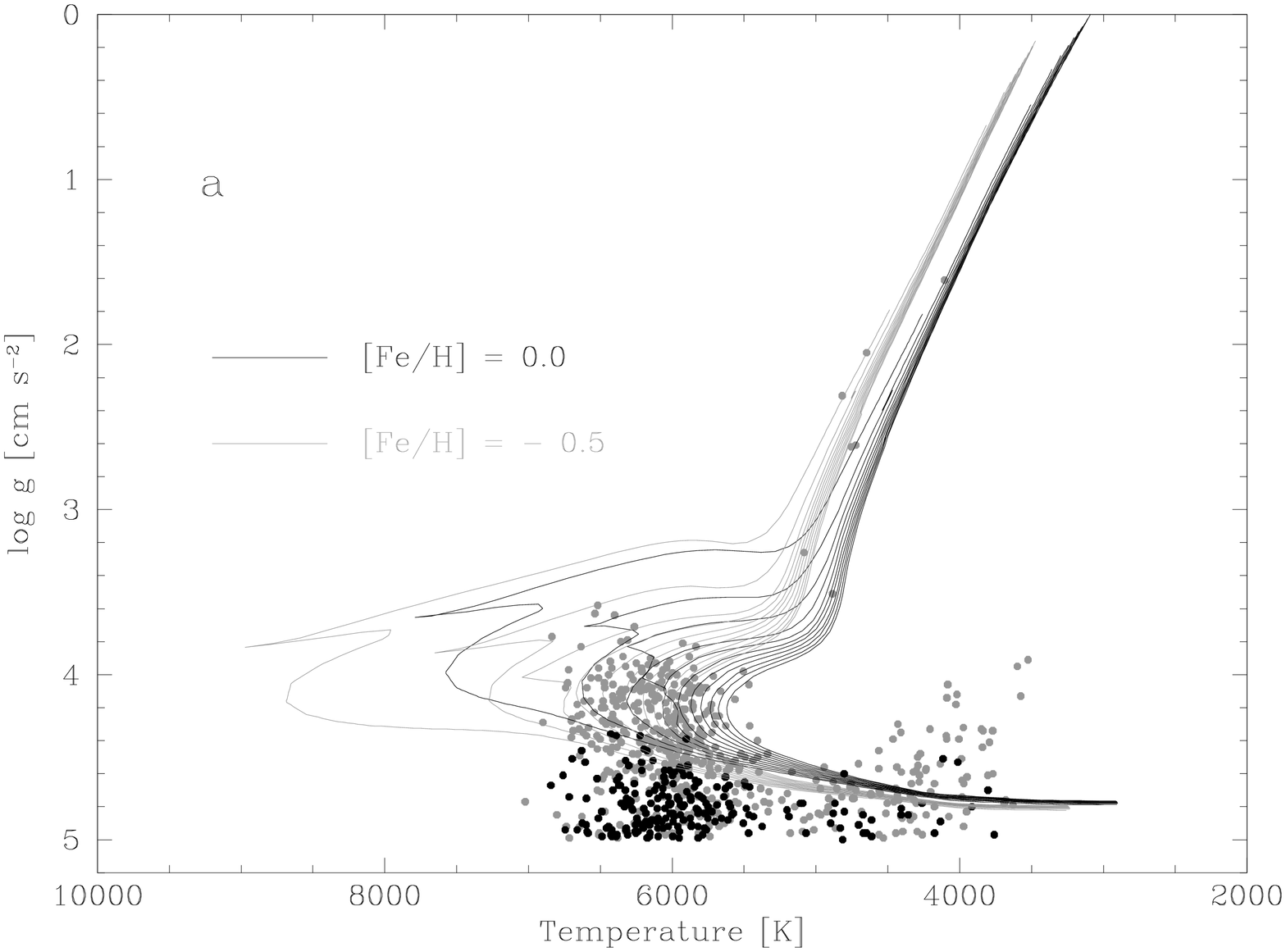} \\
\includegraphics[width=6.9cm,angle=270]{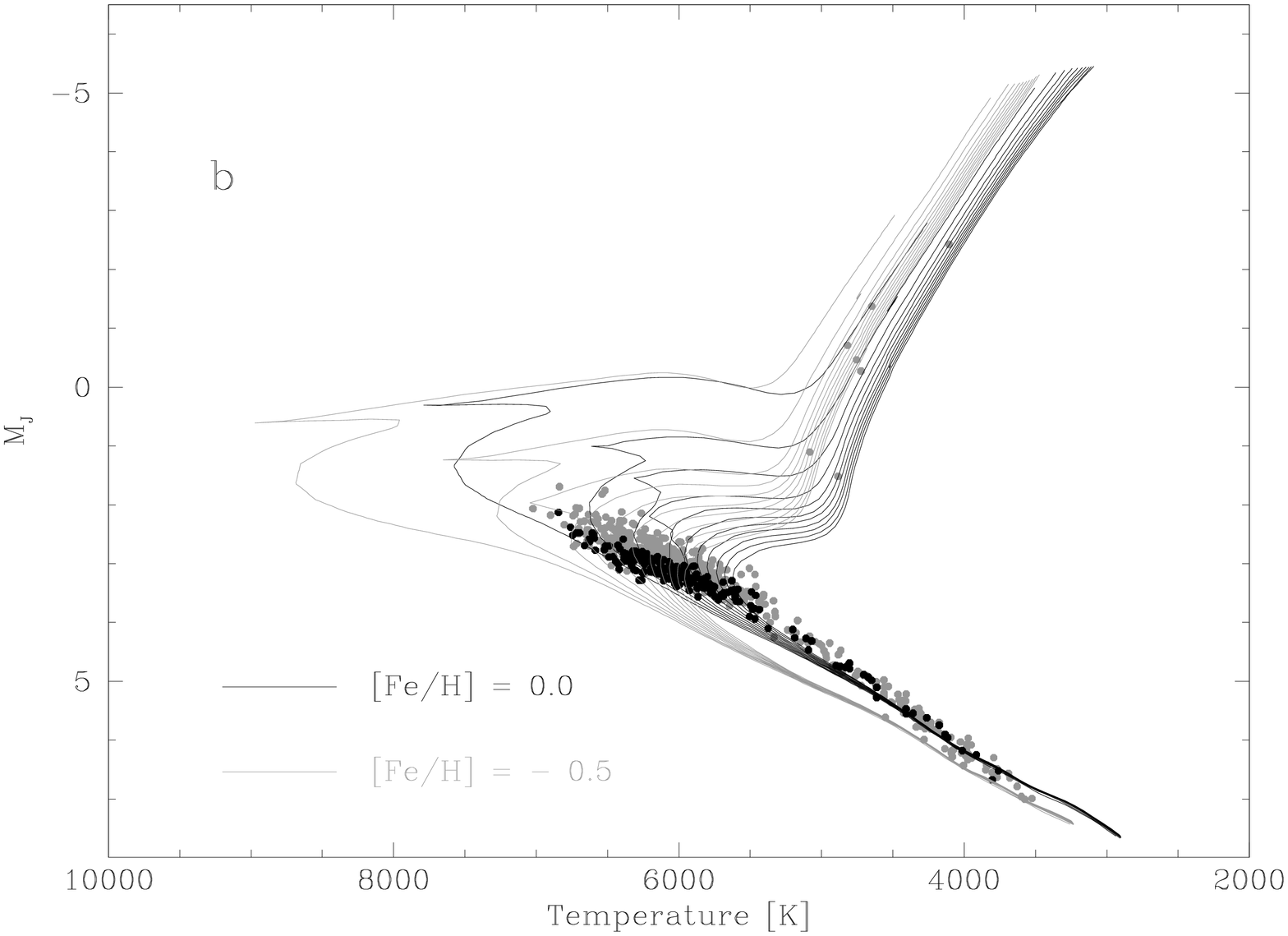}
\caption{
HR diagram of Hipparcos stars closer than 130~pc. 
The top panel (a) shows RAVE measurements of gravity and effective temperature, 
while the bottom panel (b) presents the result of our calculation of the 
absolute $J$ magnitude using the YY set of isochrones. 
Grey points are Hipparcos stars with a trigonometric distance error smaller 
than 20\%, a spectroscopic distance error below 32\%\ and with RAVE 
spectra with S$/$N~$>40$. Black points are the subset where the B10 and 
the three isochrone set calculations presented in this paper found a 
spectroscopic distance error smaller than 32\%. 
Black and gray lines  trace YY isochrones with 
Solar and subsolar composition and for ages between 1 and 11 Gyr with a 
1 Gyr spacing. The vast majority of Hipparcos stars close enough 
to have accurate trigonometric parallaxes are dwarfs.
}
\label{FIGHRHipparcos}
\end{figure}

\begin{figure}[hbtp]
\centering
\includegraphics[width=7.1cm,angle=270]{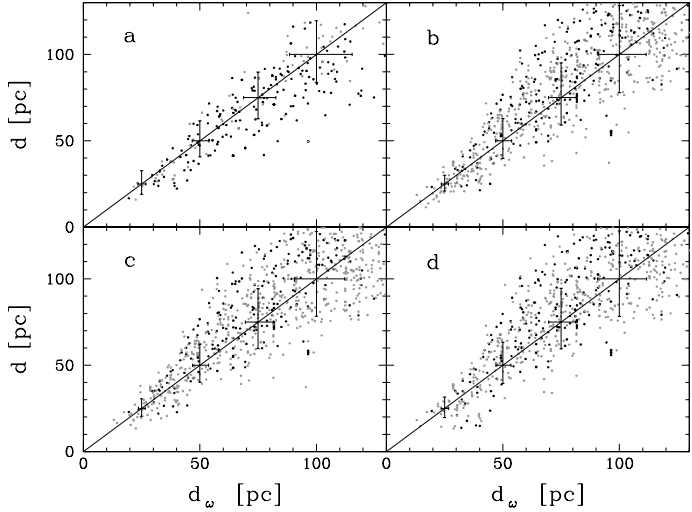}
\caption{Comparison of distances to Hipparcos stars. The trigonometric
parallax distances $d_\varpi$ are compared to spectroscopic distances 
$d$ from B10 (a), and to values derived here using the 
YY (b), Dartmouth (c), and Padova (d) isochrones. 
We are using the same 
selection criteria and color scheme as in Fig.~\ref{FIGHRHipparcos}.
The errorbars show average formal errors for stars in four distance 
bins. The diagonal line is a $1:1$ relation.
%Dataset B: all points  d_me = k * d_omega
}
\label{FIGall_hip_distances}
\end{figure}

\begin{figure}[hbtp]
\centering
\includegraphics[width=7.1cm,angle=270]{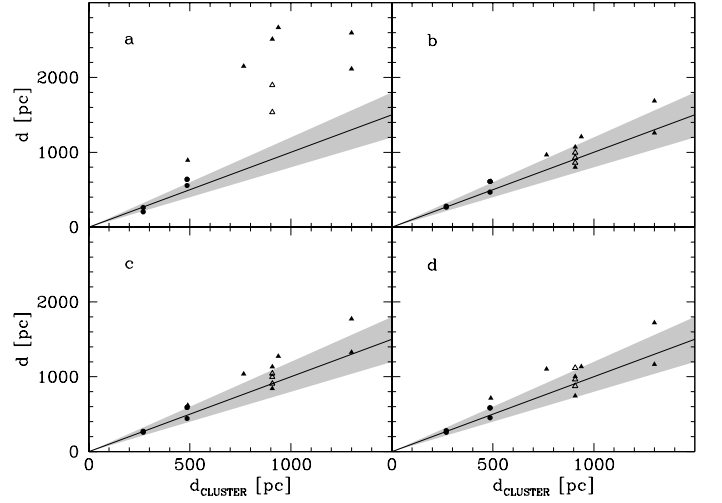}
\caption{
Distances to open clusters ($d_\mathrm{CLUSTER}$) from Dias et al.\ (2002, 2007) 
compared to spectroscopic distances $d$ of cluster members observed 
by RAVE from B10 (a), and to values derived here using the 
YY (b), Dartmouth (c), and Padova (d) isochrones. Circles 
denote dwarfs ($\log g_\mathrm{RAVE} > 3.8$) and triangles likely giants 
($\log g_\mathrm{RAVE} \le 3.8$). Open symbols mark the likely members 
of M~67.
The line marks the 1:1 relation. The pairs of distances in the shaded area 
are equal to within 20\%.
}
\label{FIGall_cluster_distances}
\end{figure}

Hipparcos stars have their distances measured through trigonometric 
parallaxes \citep{vanLeeuwenbook}. Our goal is to use objects with well 
determined trigonometric distances and with high-quality RAVE spectra. 
This choice can then allow us to check for systematic errors in our 
distance computation which is the primary goal of our test. The formal 
errors on the parallaxes are $1.0 \pm 0.5$~mas, so we adopted a lower 
limit of 7.7~mas for the parallax, corresponding to a distance 
limit of 130~pc. Moreover, for this test we considered only RAVE spectra 
with a S$/$N ratio $> 40$. RAVE observed 1525 such stars so far. 
Finally we require that a formal error in 
spectroscopic distance modulus $<0.6$~mag (corresponding to a distance 
error of 32\%). The last constraint is satisfied 
by 725 spectra for calculations based on the YY isochrones, 776 spectra 
for the Dartmouth set, and 674 for the Padova set. 277 of them 
had their spectroscopic distances already included in B10. 

RAVE confirmed that most of these stars have a Solar-like composition
([M$/$H]$ = -0.12 \pm 0.25$). Fig.~\ref{FIGHRHipparcos}a shows 
RAVE measurements of the effective temperature and surface gravity. 
As expected nearly all stars are dwarfs. The magnitude limits of the 
RAVE survey also imply that most stars are somewhat hotter (and brighter) 
than the Sun, lying in the general turn-off region. As discussed 
in Sec.~\ref{s:assumed_errors} their temperature determination is 
rather accurate, but their gravity errors are quite high. 
So RAVE is able to determine their general main-sequence type, but 
not any further details. The error on gravity is even worse for 
cool stars, so the small group of deviant stars at the 
bottom-right corner of Fig.~\ref{FIGHRHipparcos}a can easily 
be normal main-sequence stars. 

The top panel of Fig.~\ref{FIGHRHipparcos} presents the input for the 
spectroscopic distance calculation, while the bottom panel depicts its 
output. The ordinate is now the calculated absolute $J$ magnitude
resulting from the YY set of isochrones. We see that the method 
nicely corrects the too-high gravities from the top panel and 
places the sub-dwarf stars back on the main sequence. Also the cool 
stars supposedly lying above the main sequence are now nicely settled 
onto it. 

Fig.~\ref{FIGall_hip_distances} compares trigonometric 
distances derived by Hipparcos ($d_\varpi$) to the RAVE 
spectroscopic distances ($d$) as determined by B10 and by the 
present calculation using the three isochrone sets. Our restriction 
to very high-quality measurements of trigonometric parallax allows us to 
ignore any transformation bias due to the non-linear relation between 
parallax and distance. 

The spectroscopic distance determinations are generally in good agreement 
with their trigonometric parallax values. Table~\ref{t:hipparcos} reports
the best fits of the relation $d = k d_\omega$ for two samples and 
for four different sets of calculations. The slopes $k$ are 
close to unity. The B10 distances are somewhat underestimated, while 
the ones reported in this paper are slightly overestimated if compared 
to trigonometric parallax measurements. On the other hand the scatter 
in $k$ is a bit larger for measurements in this paper than was the case in 
B10. This is partly due to a difference in the assumed 
errors of stellar parameters. While B10 use the same error for all spectra, 
our errors are variable and depend on the type of star and 
the S$/$N ratio of the spectrum. Our assumed errors are generally larger,
a typical error in gravity is 0.45~dex compared to 0.3~dex used by 
B10. 
%our typical errors in logg 0.45, in T: 290K, in M/H:0.13
Assuming fixed error values from B10 (300~K in temperature and 0.3~dex 
in gravity) we get more satisfying fits for the Sample~B:
$k = 1.03 \pm 0.01, \sigma = 0.21$ (YY isochrones),
$k = 1.04 \pm 0.01, \sigma = 0.21$ (Dartmouth), and 
$k = 1.05 \pm 0.01, \sigma = 0.22$ (Padova set).

Note that many of the Hipparcos stars with the temperature close 
to 6000~K have $\log g \sim 4.8$, a value too high for this part 
of the main sequence (Fig.~\ref{FIGHRHipparcos}a). This offset 
challenges the distance routine which assumes a Gaussian 
error distribution. The offset is more frequent for 
the Hipparcos stars than for typical RAVE spectra. 
About 15\% of the Hipparcos stars hotter than 5500~K have a claimed 
$\log g > 4.8$, compared to just 8\% of the cases for the general 
RAVE sample. Also, Hipparcos stars sample the cool part of the 
main sequence down to temperatures of 3500~K. But the general 
RAVE sample has hardly any dwarfs cooler than 4500~K, with a 
vast majority at around 6000~K. The difference is due to 
a magnitude limited nature of the survey, where cool stars are 
sampled only in a small volume around the Sun. 

So the scatter obtained for the Hipparcos stars may not be 
representative of the entire population of RAVE spectra of the 
main-sequence stars. Note that the Hipparcos stars 
are generally brighter than the general sample observed by RAVE, 
so their spectra have a higher $\SN$ ratio. Thus the 
distance error estimates for the Hipparcos stars may be somewhat 
optimistic. It is interesting to note that the smallest 
distance errors are frequently found for stars with overestimated 
gravity values. This is understandable,
as for such stars the main-sequence solution is an obvious choice. On 
the other hand the stars with more realistic and so lower values of 
gravity could also be turn-off stars, so their absolute magnitude 
determination is less certain.  

To summarize, trigonometric parallax measurements and 
results of our spectroscopic distance computations show  
average discrepancies of $\simlt 5$\%. The scatter for measurements of 
individual stars is $\simlt 24$\%. Including the formal error of the 
Hipparcos parallaxes in quadrature, we find that the spectroscopic
distances are accurate to within $\simlt 22$\%.
 
\subsection{Distances to members of star clusters}
\label{s:clusters}

RAVE observed virtually no giant stars with an accurate measurement of 
their trigonometric parallax. Our spectroscopic distance 
measurement, however, can easily reach the giants that form about 
half of the RAVE sample. So we used known members of open and globular 
clusters observed by RAVE to extend our check of the distance 
scale. %serendipitously

%\begin{landscape}
\begin{table*}
%\begin{deluxetable}{lrrlrrrl@{$\pm$}r@{$\pm$}lr@{$\pm$}lr@{$\pm$}l}
%\rotate
%\tabletypesize{\tiny}
%\tablecaption{
\caption{
Distances to members of star clusters.
}
\label{t:clusters}
%\tablewidth{0pt}
%\tablecolumns{18}
\centering
{\small
\begin{tabular}{lrrlrrr@{$\pm$}lr@{$\pm$}lr@{$\pm$}lr@{$\pm$}l}
\hline\hline
%\tablehead{
Object-ID       &RA$_\mathrm{J2000.0}$&Dec$_\mathrm{J2000.0}$&\multicolumn{2}{c}{Cluster}&S$/$N&\multicolumn{8}{c}{Spectroscopic Distances from RAVE [pc]}\\
                & [deg]        & [deg]         &ID & d [pc]                &     &\multicolumn{2}{c}{B10}&\multicolumn{2}{c}{Yonsei-Yale}&\multicolumn{2}{c}{Dartmouth}&\multicolumn{2}{c}{Padova}\\
%}
\hline
%\startdata
%# OBJECTID        RA            DE           cluster clusterdist S2N  Breddist               YYdist           Dartdist          Padovadist %Padova using K mag
%#
J000128.6-301221 &   0.369417& --30.205861&  Blanco 1 &    270 &   64  &    260 &  80             &  270&  70 &  260 & 60 & 260 & 60\\ %258  
J000324.3-294849 &   0.851583& --29.813722&  Blanco 1 &    270 &   64  &    200 &  30             &  280&  50 &  270 & 50 & 280 & 50\\ %270  
OCL00148\_1373319& 114.302500& --13.748667&  NGC 2423 &    770 &   73  &   2150 & 960             &  960& 220 & 1040 &220 &1100 &190\\ %1061 
OCL00147\_1373471& 114.434792& --14.837000&  NGC 2422 &    490 &   44  &    890 & 400             &  610&  90 &  620 & 90 & 710 &160\\ %683  
J075214.8-383848 & 118.061750& --38.646889&  NGC 2477 &   1300 &   50  &   2110 &1200             & 1280& 310 & 1330 &310 &1160 &320\\ %813  
J075242.7-382906 & 118.178125& --38.485083&  NGC 2477 &   1300 &   61  &   2600 &1300             & 1690& 320 & 1770 &320 &1720 &340\\ %1687 
        M67-0105 & 132.821250&   11.804444&  NGC 2682 &    910 &   89  &\multicolumn{2}{c}{\ldots}&  800& 170 &  840 &170 & 740 &150\\ %606  
     M67-0135(*) & 132.839833&   11.768361&  NGC 2682 &    910 &   50  &\multicolumn{2}{c}{\ldots}&  930& 280 & 1000 &290 &1120 &280\\ %1074
        M67-0223 & 132.932875&   11.945083&  NGC 2682 &    910 &   57  &   2510 & 700             & 1070& 260 & 1130 &270 &1010 &250\\ %823  
     M67-2152(*) & 133.045750&   11.530361&  NGC 2682 &    910 &   48  &   1900 & 940             & 1000& 270 & 1040 &270 & 960 &200\\ %857
     M67-6515(*) & 133.069083&   11.327361&  NGC 2682 &    910 &   99  &   1540 & 770             &  860& 170 &  910 &170 & 880 &190\\ %658
OCL00277\_2236411& 165.825000& --58.488444&  NGC 3532 &    490 &   52  &    640 & 270             &  610& 120 &  590 &120 & 580 &120\\ %576  
OCL00277\_2236511& 165.934708& --58.960972&  NGC 3532 &    490 &   39  &    560 & 250             &  470& 150 &  440 &130 & 450 &150\\ %443  
  T7751\_00502\_1& 171.452250& --43.164611&  NGC 3680 &    938 &   67  &  2670  & 990             & 1210& 280 & 1270 &280 &1140 &260\\ %938
J125905.2-705454 & 194.771667& --70.914972&  NGC 4833 &   5900 &   37  &\multicolumn{2}{c}{\ldots}& 5500&1000 & 5900 &1100&4900 &1100\\%4350 
\hline
% OCL00404_1786252 Trumpler 28 RV too much different!  
%\enddata
\end{tabular}
}
\tablefoot{(*) A likely cluster member: Simbad does not list it as a cluster member, but it is within the cluster area and has  
a radial velocity consistent with the cluster membership.}
%\end{deluxetable}
\end{table*}
%\end{landscape}

The Vizier catalog of open clusters of \citet{dias02}, Version~2.8 was used. 
A star is considered to be a known open cluster member if: 
(i) it lies within the apparent diameter of the cluster center in both 
coordinates, (ii) its proper motion and the value for the cluster 
reported by \citet{dias02} do not differ by more than the
sum of their reported uncertainties,
(iii) the same is true also for radial velocities  
\citep[if][do not report the radial velocity error a difference of 2 \kms\  
is considered acceptable]{dias02}, 
(iv) the Simbad catalog reports that the object is a ''star in cluster'' 
(class *iC). Clearly these requirements are conservative. They 
exclude many potential cluster members observed by RAVE. But our goal 
here is to use definite cluster membership to check our distances.  
We also checked the 
\citet{harris96} catalog of globular clusters. The star 
J125905.2-705454 is 3.43~arcmin from the center of the globular cluster 
NGC~4833, which has a half-mass radius of 2.41~arcmin. Its heliocentric 
radial velocity of $202.1 \pm 0.8$~\kms\ (as measured by RAVE) matches the 
cluster's radial velocity of $200.2 \pm 1.2$~\kms. Similarly its
RAVE metallicity [M$/$H]$ = -1.2$ is consistent with the cluster 
iron abundance of [Fe$/$H]$ = -1.79$. So we consider it to be a 
member of this globular cluster. No other members of globular clusters 
were found. 

Fig.~\ref{FIGall_cluster_distances} and Table~\ref{t:clusters} summarize 
the results. Altogether we have observations of 11 secure members of 
7 different open clusters, one globular cluster member and 3 additional 
likely cluster members. 
Four stars are dwarfs ($\log g > 3.8$), the rest are giants ($\log g \le 3.8$). 
Our results show a satisfactory 
performance for dwarfs and for giants. Globular cluster member J125905.2-705454 is 
omitted from Fig.~\ref{FIGall_cluster_distances} due to scale. 
Table~\ref{t:clusters} shows that also for this supergiant 
($\teff = 4250$~K, $\log g = 0.8$) the spectroscopic distance is 
consistent with the value from the literature. The derived spectroscopic 
distances are generally within 20\% of the distance to the cluster as 
given in \citet{dias02} and \citet{harris96}. The average and the scatter of the 
spectroscopic to cluster distance ratio for the entries in Table~\ref{t:clusters} equals 
$0.97 \pm 0.14$ ($1.11 \pm 0.08$), 
$1.03 \pm 0.15$ ($1.16 \pm 0.08$), and 
$0.89 \pm 0.15$ ($1.11 \pm 0.08$)
if using the YY, Dartmouth, and Padova isochrones respectively. The values 
in brackets are obtained if we fit only the open cluster members. The difference 
of solutions inside and outside the brackets is quite large, given that  
it corresponds to a contribution of a single globular cluster member.
So we note that all values are consistent with the 1:1 relation  
to within $\sim 10$\%.

B10 use observations of the open cluster NGC~2682 (M~67) to study the distance 
scale of giants. The distances found are generally 
too large. Here we show that our distances are consistent with the 
adopted value of 910~pc from the literature. The two members in 
Table~\ref{t:clusters} give $930 \pm 160$~pc (YY isochrones), 
$980 \pm 160$~pc (Dartmouth), and $880 \pm 150$~pc (Padova). If we 
would add the likely M~67 members by their heliocentric radial velocity 
($31.2$~\kms$ < RV < 33.4$~\kms) the number of observed M~67 members would 
increase to five and the results would be 
$930 \pm 110$~pc (YY), $980 \pm 110$~pc (Dartmouth),
and $940 \pm 100$~pc (Padova). 

Stellar clusters have a known value of the color excess $E_{B-V}$, so they
allow some discussion of the influence of interstellar extinction on our 
distance estimates. 
\citet{munariadpspaper2} give for the extinction in the $J$ band 
$A_J \simeq 0.887  E_{B-V}$. Nearly all clusters in 
Table~\ref{t:clusters} have $E_{B-V} < 0.1$, so their extinction in 
the $J$ band is $A_J < 0.089$. Ignorance of extinction makes their 
distances overestimated by no more than 4\%. The only clusters 
with the color excess 
larger than 0.1 are NGC~2477 \citep[$E_{B-V} = 0.24$, ][]{dias02} 
and NGC~4833 \citep[$E_{B-V} = 0.33$, ][]{harris96}. Neglecting  
interstellar extinction should thus make their distances overestimated 
by 10\% and 14\%, respectively. Note that these two clusters lie very 
close to the Galactic plane ($b = -5.8^\circ$ and $b = -8.0^\circ$, respectively).
A vast majority of the RAVE sample has $|b| > 20^\circ$, so its  
interstellar extinction can be safely ignored.

%dist cluster_num cluster_name E(B-V)
%1100 414 Collinder 121 0.04
% 760 550 NGC 2423      0.097 (already in table 1 but has many more members)
% 800 437 ASCC 33       0.05
% 400 887 Alessi 5      0.15
RAVE observed more objects that are likely members of open clusters and 
satisfy the same selection criteria, with the difference that Simbad does
not recognize them as cluster members. There are a number of such likely 
members of NGC~2423, and a few objects in the open clusters Collinder~121, 
ASCC~33, and Alessi~5. Here we compared spectroscopic distance measurement 
of confirmed cluster members to known cluster distances. A study of 
possible cluster members and applications using spectroscopic distances to 
stellar clusters will be published elsewhere.

\section{Derived distances}
\label{s:results}

\begin{figure*}[hbtp]
\centering
\includegraphics[width=\textwidth,angle=0]{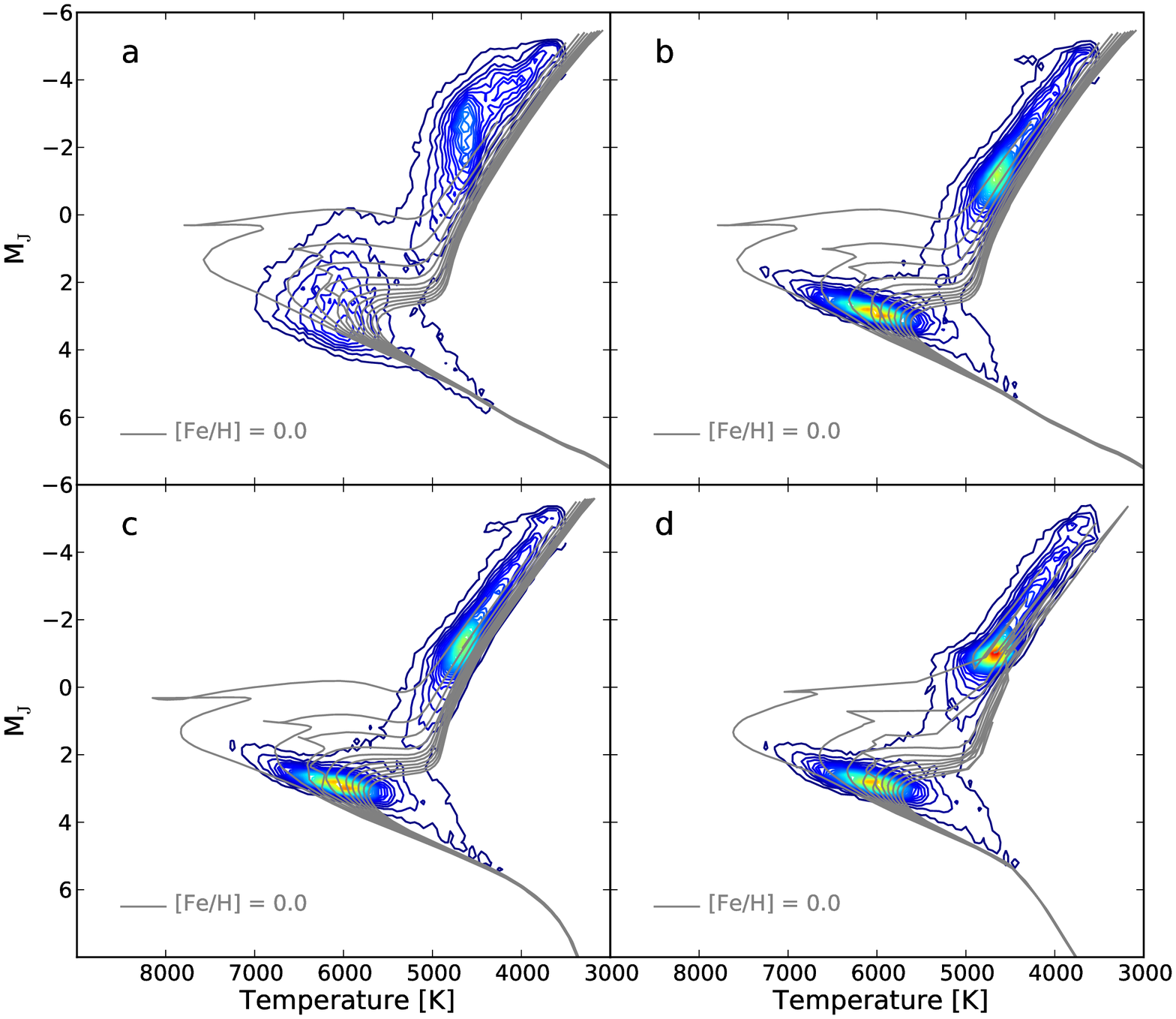}
\caption{
Frequency of occurence  of 
solutions  in  the  absolute  J  magnitude  vs. 
temperature  diagram  for  four sets of calculations: 
(a) results from the B10 paper, (b-d) new calculations using 
the YY, Dartmouth and Padova isochrones, respectively.
All panels include the same set of $145\,028$ spectra common to all 
four calculations, but the appearance of (b-d) for the whole dataset 
of $235\,064$  stars would be almost identical. 
Fifty colored iso­-frequency  contours  are 
linearly spaced between 20 and 1900 stars per bin with a step 
of 40. Each bin spans 50~K in the temperature and 0.2~mag in 
the magnitude direction. 
Gray lines  trace  Solar  composition  
isochrones  between  1 and 11 Gyr with a 1 Gyr spacing.
}
\label{FIGtest106}
\end{figure*}

The method described in Sec.~\ref{s:method} has been applied to a 
sample of RAVE stars with spectroscopically determined values of 
temperature, gravity and metallicity. First we describe the sample 
and stellar properties as reflected in the derived absolute magnitudes.
Then we move to statistical properties of derived distance distributions, 
including their error estimates. We conclude with a detailed discussion 
of the catalog and its availability. 

\subsection{The sample}

RAVE is an ongoing survey that uses the multi-fiber spectrograph of the UK
Schmidt telescope at the Anglo Australian Observatory to obtain stellar
spectra with a resolving power of $\sim 7500$ in the wavelength range from
8410~\AA\ to 8795~\AA. It is conducted by an international collaboration led
by M.\ Steinmetz and aims to obtain a million stellar spectra in the
magnitude range $9 < I_{\rm DENIS} < 13$. The mode of the signal-to-noise
ratio per pixel of the $>400\,000$ that have so far been gathered follows the
relation $\SN = 10^{-0.2(I_{\rm DENIS} - 19.1)}$ (Z08).  A detailed
description of the survey is given in the papers accompanying the first two
public data releases \citep{S06,Z08}, soon to be followed by the third one
\citep{siebert10}.

This paper is based on the sample of $332\,747$ currently 
reduced spectra with a measurement of radial velocity. For $292\,261$
of these we have determined the values of stellar parameters, with the 
rest being either too noisy or suffering from systematic problems such 
as a contamination by the second interference order. A subsample 
of $251\,364$ have $\SN > 20$, so  dependable estimates of the 
values of stellar parameters (temperature, gravity and metallicity),
which are virtually free from systematic offsets, could be obtained
(Z08). The S$/$N ratio depends only on the $I_{\rm DENIS}$ magnitude 
of the object, on the observing conditions, and on the performance of the 
instrument.  Objects to be observed are magnitude selected and 
observed in a random order. So our decision to discuss only objects with 
$\SN > 20$ is equivalent to a modified magnitude cut and can be 
taken into account when discussing statistical properties of the sample.   
A few percent of the stars turn out to be double-lined spectroscopic 
binaries, coronal emission objects, or show other peculiarities. Our 
distance estimates assume the spectrum belongs to a single normal  
star. So these peculiar objects need to be excluded, for which we used 
an automated classification routine \citep{matijevic10}. 
The routine unavoidably fails to identify some of the binary stars. 
In the case of a single--lined binary star the distance would be unaffected.
On the other hand the distances to unresolved double--lined spectroscopic 
binaries are underestimated by up to a factor of $\sqrt{2}$. But such
worst cases are rare \citep{matijevic10}. If a star in a binary 
is two magnitudes brighter than its unidentified companion, the error 
in its estimated distance drops to 8\%. The final sample of stars
for which the distances have been estimated contains $235\,064$ 
spectra with $\SN > 20$ and without detected peculiarities.
They belong to $210\,872$ different stars.

\subsection{Derived absolute magnitudes}

The immediate result of our method is an estimate of the absolute 
magnitude of the star. We use the $J_{\rm2MASS}$ passband because 
the apparent magnitudes measured by the 2MASS survey are accurate, 
available for virtually all stars in our sample (Z08, Table~7)  
and sufficiently red to allow us to ignore any effects of interstellar 
extinction (see Sec.~\ref{s:clusters}). 

The properties of the sample can be studied with the H-R diagram 
in Fig.~\ref{FIGtest106}, which shows the results of the absolute 
magnitude calculations of B10 and by the method presented 
here for all three isochrone sets. The relative density of stars is 
represented by linearly spaced colored contours. A subset of 
Solar metallicity isochrones is overlaid for illustration purposes.

Two groups of stars are apparent: 
the moderate temperature stars which are close to the main sequence 
and the group of cool giants. Cool dwarfs are almost missing. This can 
be understood, because the RAVE survey has a fairly limited magnitude 
span ($9 < I_{\rm DENIS} < 13$), so the volume of space with cool dwarfs 
of appropriate apparent magnitude is smaller than the one with hotter 
dwarfs and much smaller than the volume occupied by luminous and 
thus distant giants. Hot dwarfs are also infrequent, which is not 
surprising given the low probability of their formation and their 
short lives. Similar selection effects diminish also the relative 
frequency of supergiants. As a result of these considerations RAVE 
observes mostly dwarfs slightly hotter than the Sun and giants with 
surface temperatures of $\sim 4700$~K.

Figure~\ref{FIGtest106}a leaves some room for improvement. 
Many of the main-sequence stars 
seem to be just leaving the main sequence, which is unlikely. Note that 
this is even more true if their metallicities were lower than Solar. 
Also, giants lie  above the Solar-metallicity isochrones. 
This would be fine if they were very metal-poor, but 
RAVE metallicities tell that they are not. Consequently B10 find that 
the distance to M67 giants is 1480 pc (if excluding the red clump stars), 
a value 60\% larger than in the literature. The B10 distances to 
Hipparcos dwarfs however turn out to be in agreement with astrometric 
distance determination of \citet{vanLeeuwenbook}.

Results using the refined method are given in Fig.~\ref{FIGtest106}b-d. 
These panels show the results for the sample of $145\,028$ 
normal spectra with $\SN > 20$ for which   B10 also give distances 
(Fig.~\ref{FIGtest106}a). The whole sample we present 
in our Catalog (Sec.~\ref{s:catalog}) is larger and includes 
$235\,064$ spectra. Both samples are drawn from the same 
selection strategy and observation protocol, implying very similar 
errors in derived values of stellar parameters and thus distances.
Results of the refined computations show that dwarf stars 
now lie very close to the main sequence and the giants are in the area 
of moderately metal-poor giants with an age of a few Gyr. 
A closer inspection reveals that most main-sequence stars 
are more luminous than the locus of the zero-age main sequence. This 
is understandable, as a main-sequence star with $\sim 6000$~K 
and an age of $\sim 5$~Gyr is close to the turn-off point
(see the 5~Gyr isochrone on Fig.~\ref{FIGtest106}),  
so it is notably brighter than it was when younger. 
 %The vertical 
%line at $\sim 3600$~K is an artefact due to 
%a lower temperature boundary in the grid of the 
%Kurucz stellar models \citep{munari05} used to estimate stellar parameter 
%values, which do not extend below $\teff = 3500$~K. 

Results for the calculations using the YY (Fig.~\ref{FIGtest106}b) and 
Dartmouth (Fig.~\ref{FIGtest106}c) isochrones are very 
similar. However, the graph based on Padova isochrones (Fig.~\ref{FIGtest106}d) 
shows a notable concentration of giants at $M_J \simeq -0.9$. 
This difference arises because only Padova isochrones include a proper
treatment of the red-clump region. Since stars  spend a long time in this
region we expect many of the giants observed by RAVE to lie in the red clump.
Hence, distances based 
on the Padova isochrones are to be preferred over ones calculated using 
the YY or Dartmouth isochrones, at least for giants. 
The locus of the red-clump stars lies at a bit hotter temperatures 
than the RGB stars with the same absolute magnitude. So the red-clump 
stars in Fig.~\ref{FIGtest106}d are not superimposed on the 
red-giant branch.

\begin{figure}[hbtp]
\centering
\includegraphics[width=7cm,angle=270]{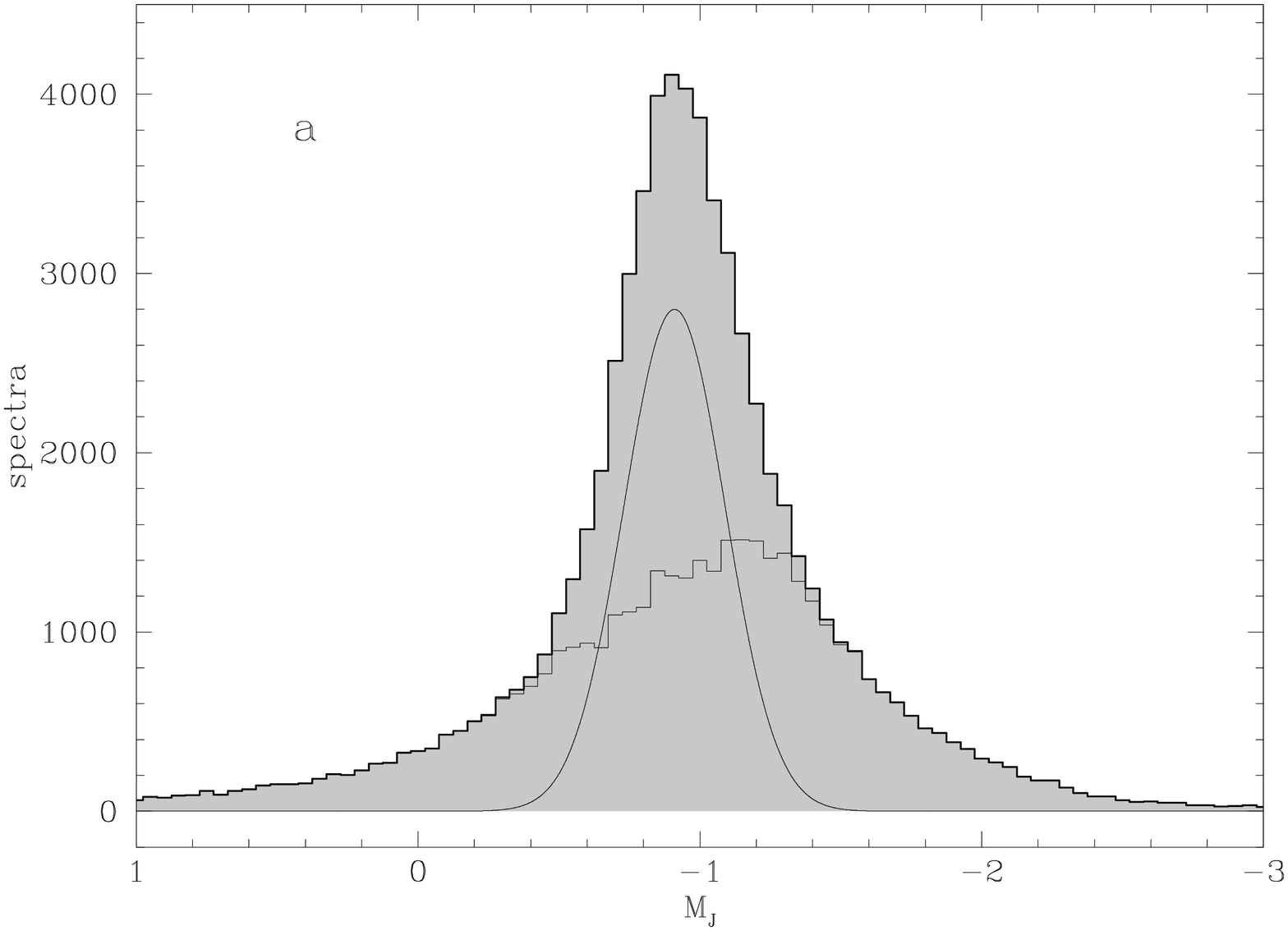}\\
\includegraphics[width=7cm,angle=270]{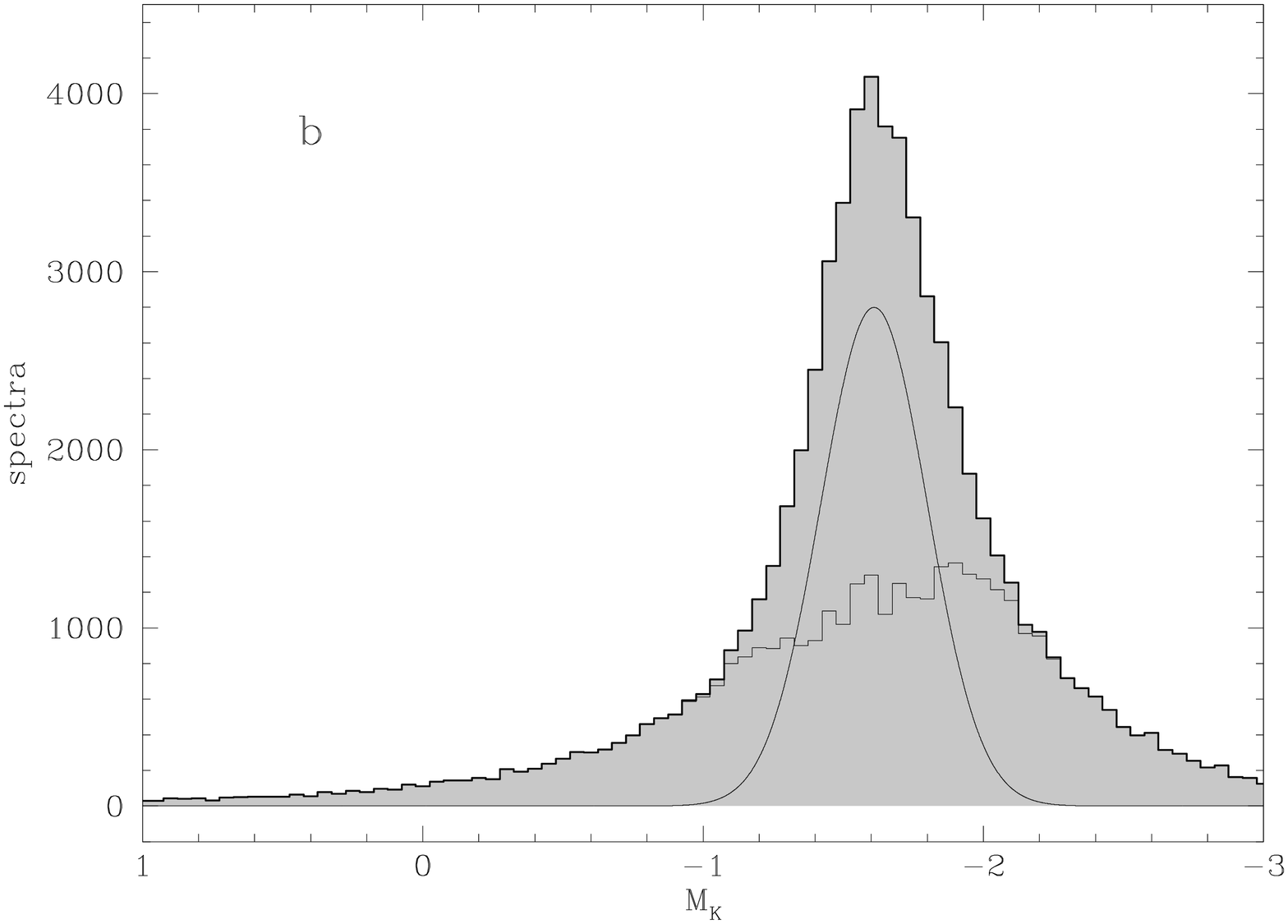}
\caption{
Histogram of absolute J (panel a) and K (panel b) magnitudes for red clump stars. 
2MASS J and K apparent magnitudes were combined with the distance 
calculation using the Padova isochrone set to derive the absolute J and 
K magnitudes. Red clump stars are assumed to have
4500~K~$<\teff <$~5000~K. So the top panel presents a 
histogram of $M_J$ values in Fig.~\ref{FIGtest106}d between 
these two temperatures. The histogram representing all stars 
in this temperature interval is decomposed into an assummed 
Gaussian red clump peak and a smooth red giant branch contribution. 
For parameter values see text.
%Siebert uses M_K=-1.6+-0.03 (see also Alves 2000)
}
\label{FIGRC_absmagnitude}
\end{figure}

The red clump region in Fig.~\ref{FIGtest106}d can be studied further. 
Fig.~\ref{FIGRC_absmagnitude}a shows a histogram of $M_{J}$ for temperatures 
between 4500~K and 5000~K. Note the pronounced red clump maximum, 
modelled as a Gaussian with the peak at $M_J = -0.91 \pm 0.02$ 
and $\sigma = 0.18$. The parameters were adjusted so that the 
underlying contribution of the RGB stars remains a smooth function 
of the absolute magnitude. RGB stars contribute 49\% of all RAVE spectra
with  $-1.5 < M_J < -0.5$ and 4500~K$< \teff < 5000$~K. If the 
absolute magnitude interval is broadened to $-3 < M_J < 1$ the
fraction rises to 62\%. So their contribution should be accounted for 
when selecting the red clump stars by their effective temperature, colour 
or absolute magnitude.  

The 2MASS survey reports also the $K$ magnitudes 
for the same stars. The histogram using the same selection criteria 
(Fig.~\ref{FIGRC_absmagnitude}b) can again be deconvolved into a 
smooth RGB contribution and a Gaussian red clump maximum with 
the peak at $M_K = -1.61 \pm 0.02$ and $\sigma = 0.19$. RGB 
stars contribute 45\% of all RAVE spectra
with  $-2.1 < M_K < -1.1$ and 4500~K$< \teff < 5000$~K. If the 
absolute magnitude interval is broadened to $-3 < M_K < 1$ the
fraction rises to 60\%.

The position of the red clump peak is in agreement with  
the values frequently used in the literature. \citet{valentini10}
used a carefully selected sample of red clump stars to estimate 
$M_K = -1.55$ and $M_J = -0.87$. Similarly \citet{vanhelshoecht07} 
give $M_K = -1.57 \pm 0.05$ and \citet{groenewegen08} derives 
$M_K = -1.54 \pm 0.04$.

\subsection{Formal errors of derived distances}

\label{s:formal_errors}
The errors in our distances arise 
from four sources. First, the values of temperature, gravity and metallicity 
as derived by RAVE have their uncertainties 
(see Sec.~\ref{s:assumed_errors}). Moreover, the isochrone solution 
can be uncertain due to intersections of isochrones in the parameter 
space. Also the isochrone positions themselves depend on assumptions 
used in their calculation. Finally, we convert the absolute $M_J$ 
magnitude into distance assuming a negligible interstellar absorption 
and a reliable measurement of the apparent magnitude by the 2MASS survey.
The last point has been already discussed and generally does not 
present a major concern. We tried to compensate for isochrone differences 
by performing all calculations three times, each time with a different 
set of isochrones. The uniqueness of the solution has been discussed 
in detail by B10. This leaves us with studying the influence of the  
errors of measured stellar parameters on the derived distances.

\begin{figure}[hbtp]
\centering
\includegraphics[width=7.1cm,angle=270]{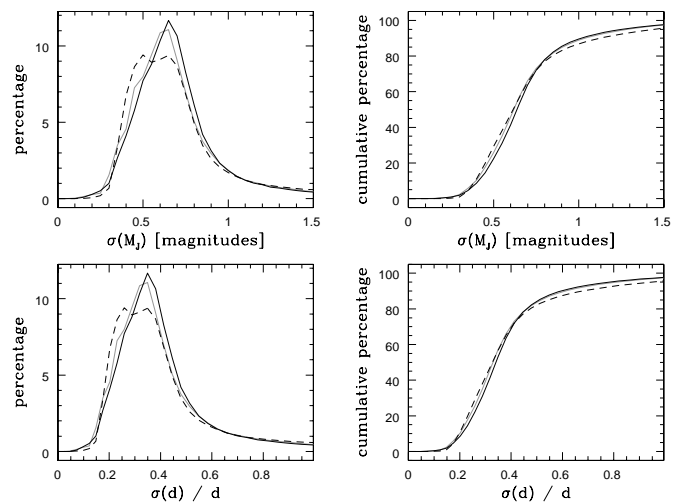}
\caption{Formal error distribution histogram (left) and cumulative plot 
(right) for absolute $J$ magnitude (top) and distance (bottom). Black
lines are for YY isochrones, grey ones for Dartmouth, and the dashed ones 
for the Padova ones. The histograms have bin widths of $0.05$ mag 
and include $235\,064$ spectra that have $\SN>20$ and do
not show any peculiarities.
}
\label{FIGerror_dist_modulus}
\end{figure}

Fig.~\ref{FIGerror_dist_modulus} plots the formal errors of the 
absolute magnitude and distance, both as a histogram and as a cumulative
plot. The results are very similar when using YY and Dartmouth isochrones,
while the Padova set yields marginally smaller errors. The reason is a 
stronger concentration of stars in the red clump region. 
%But note 
%that for stars with an absolute magnitude $M_J \simeq 1$, caught 
%in-between the main sequence and red clump region 
%(see Fig.~\ref{FIGtest106}d) the calculations using the Padova set 
%cannot easily decide 
%whether to push them down towards the main sequence or up towards 
%the red clump region. So these stars, which comprise $\sim 5$\%
%of the whole sample have very large distance errors. The 
%calculation based on Padova isochrones therefore yields overall 
%lower errors, except for stars with gravity values that place them 
%in the region between the main sequence and the giant region. 
 There are virtually no stars with distance errors smaller than 10\%, 
which is understandable. On the other hand half of the stars have their
distances determined to better than 33\%. These results 
are based on the formal errors obtained from 
eq.~\ref{eq:sigma} and the errors in the parameter described in 
Sec.~\ref{s:assumed_errors}.

\begin{figure}[hbtp]
\centering
\includegraphics[width=8.9cm,angle=0]{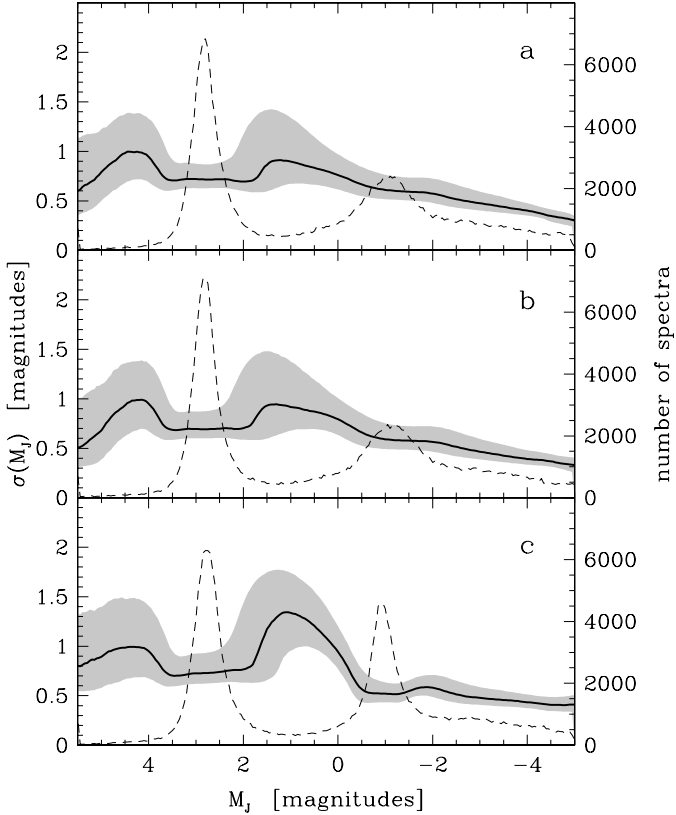}
\caption{Error in absolute magnitude as a function of 
absolute $J$ magnitude. The thick lines show running median errors for 
calculations using YY (a), Dartmouth (b) and 
Padova isochrones (c). The Boxcar for the
running median is $\pm 0.5$ magnitudes. The shaded area shows the 20 to 80
percentile range of the error distribution and the dashed line traces the
number of objects per $0.1$ mag wide absolute magnitude bin.
}
\label{FIGerror_scatter_in_HR}
\end{figure}

\begin{figure}[hbtp]
\centering
\includegraphics[width=10cm,angle=0]{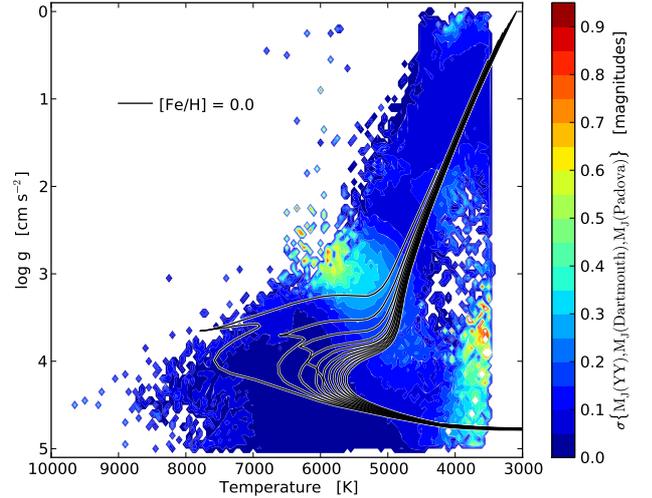}
\caption{Comparison of absolute $J$ magnitudes derived using either 
YY, Dartmouth, or Padova isochrone sets. Colors label an average
dispersion in magnitudes for non-peculiar spectra with $\SN > 20$ 
in a given (temperature,~gravity) bin. Solar composition isochrones 
from the YY set with ages between 1 and 11 Gyr and with a 1 Gyr 
spacing are overlayed. Metal-poor isochrone sets would be shifted 
towards higher temperatures (see Fig.~\ref{FIGHRHipparcos}a).
}
\label{FIGdistance_comparison}
\end{figure}

The errors of the stellar parameters 
vary across 
the H-R diagram, so it should be the same also with the errors of 
derived distances. The specific structure of the RAVE sample, which 
has a negligible number of very young stars still settling on the 
main sequence and which lacks large numbers of high-mass main-sequence 
stars, implies that the absolute magnitude $M_J$ could be used to 
label the position of the star on the H-R diagram. In effect this 
means that we can use the vertical coordinate in Fig.~\ref{FIGtest106}
as an independent coordinate for studying distance errors over the 
H-R diagram.
Fig.~\ref{FIGerror_scatter_in_HR} plots the formal scatter of the 
absolute $J$ magnitude for different luminosity classes. The thick 
line traces the median error and the shaded area marks the 
20\% to 80\% percentiles of the error distribution. The dashed 
line shows the relative number of objects per absolute magnitude 
bin. It is encouraging to note that the distance errors 
are lowest for stars on the main sequence and in the red clump, i.e.\ 
the most frequently occurring objects in our sample.
On the other hand 
the errors are the largest for stars with $M_J \sim 1.0$, i.e.\ 
for stars that are just leaving the main sequence. This is 
understandable as this is the region in which isochrones of different 
ages intersect in the (temperature, gravity) plane, making it
possible to find a wide variety of possible solutions. Another 
region with large uncertainties is that occupied by cool main-sequence stars.  
Here the errors are large on account of uncertain values of the gravity from
RAVE (see Sec.~\ref{s:assumed_errors}). 
For some of these stars
RAVE measures  gravities that are too low, so the stars are placed 
above the main sequence 
(see Fig.~\ref{FIGHRHipparcos}). Thus both the main-sequence and 
giant solutions enter eq.~\ref{eq:sigma} and contribute to an 
increased scatter. Fortunately these regions of increased
distance scatter are sparsely populated in our sample.

Stellar distances are derived using either of the three isochrone 
sets. These sets are not identical, so it is interesting to 
compare the results. Fig.~\ref{FIGdistance_comparison} shows the  
scatter of the three absolute magnitude values derived for each spectrum. 
Absolute magnitudes for the 
upper main sequence ($\teff \simeq 6500$~K, $\log g \simeq 4.0$) 
and red clump ($\teff \simeq 4750$~K, $\log g \simeq 2.5$) 
are almost identical for all  isochrone sets.
At the same time these regions have the largest number of stars in 
our sample (Fig.~\ref{FIGtest106}). On the other hand the most discordant results are for 
stars in the transition 
zone between the main sequence and giants. Here some isochrone sets 
favour the main-sequence and others the giant solution. The 
dispersion is somewhat elevated also for stars more luminous than 
the red clump ($\log g \simeq 1.9$). The origin of this discrepancy lies in 
differences 
in the physics included in computations of individual isochrone sets. 
Only the Padova set allows for the red clump, so the distances 
computed with this set are to be preferred for giants.

\subsection{Errors of derived distances from repeated observations}

Most of the RAVE objects are observed only once. But $\sim 10$\% 
of the observing time is devoted to repeated observations.  
This allows another check of the derived distance errors. Multiple 
observations of the same object are treated independently in our 
calculations, so the results can be directly compared. Note 
that repeated observations allow for a very efficient testing of the 
distance scatter, but they cannot point to possible systematic offsets of 
the mean. We believe that tests using Hipparcos stars with known 
parallaxes and observations of stars in clusters (Sec.~\ref{s:tests})
give sufficient confidence that any such offsets are insignificant for 
our results. 

\begin{figure}[hbtp]
\centering
\includegraphics[width=7.1cm,angle=270]{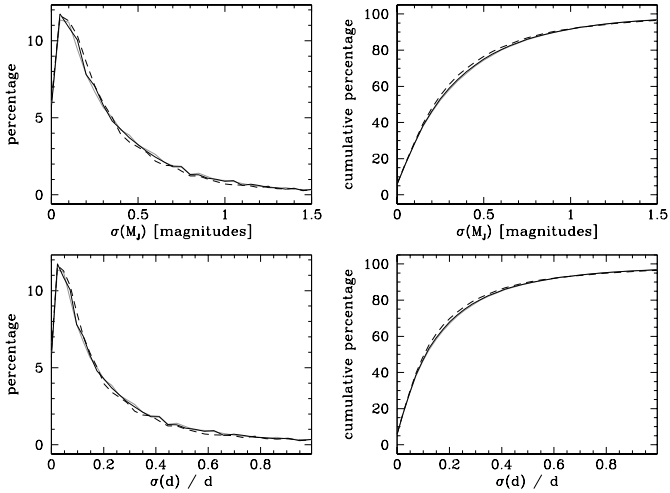}
\caption{Scatter of repeated observations of the same object. 
Scatter distribution histogram (left) and cumulative plot 
(right) for absolute $J$ magnitude (top) and distance (bottom). 
The number of repeated objects with distances calculated 
with a given set of isochrones is $19\,714$. %was $12\,847$. %12.843 (Dartmouth) 
The presentation follows the scheme of Fig.~\ref{FIGerror_dist_modulus}.
}
\label{FIGrepeat_error_dist_modulus}
\end{figure}

The total number of repeated objects in our sample is $19\,714$ for 
which $50\,149$ spectra have been obtained.  This 
list is limited to objects with identical 2MASS identifiers 
for which RAVE has at least two spectra with $\SN > 20$ and 
which do not show any 
peculiarities. Fig.~\ref{FIGrepeat_error_dist_modulus} plots the 
scatter of repeated observations in absolute magnitude and 
distance. The presentation follows that of 
Fig.~\ref{FIGerror_dist_modulus}. 
The results from repeated observations point to a remarkable 
repeatability of our distance estimates. Half of the objects 
show a distance scatter of less than $\simeq 11$\%, and 
in 80\% of the cases the distance 
scatter is smaller than $\simeq 31$\%. Repeated observations 
have a similar distribution in $\SN$ as the other observations, 
with values which are only slightly higher. 
So Fig.~\ref{FIGerror_dist_modulus} is representative also of  
the formal errors of repeated observations.  Comparison of 
Fig.~\ref{FIGerror_dist_modulus} and Fig.~\ref{FIGrepeat_error_dist_modulus}
suggests that our formal error estimates are rather conservative.

\begin{figure}[hbtp]
\centering
\includegraphics[width=8.9cm,angle=0]{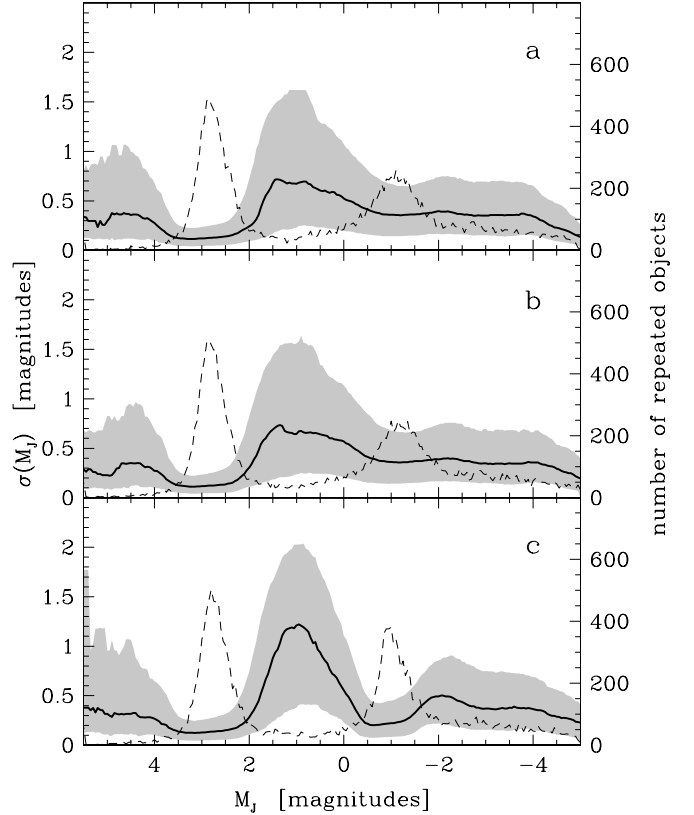}
\caption{Scatter of absolute magnitudes calculated for repeated observations 
of the same object as a function of 
absolute $J$ magnitude. Thick lines show running median errors for 
calculations using YY (a), Dartmouth (b) and Padova isochrones (c). 
The Boxcar for the running median is $\pm0.5$ magnitudes. The shaded 
area shows the 20 to 80 percentile range of the error distribution and 
the dashed line traces the number of repeated objects per $0.1$ mag wide 
absolute magnitude bin.
}
\label{FIGrepeat_error_scatter_in_HR}
\end{figure}

The scatter of distances derived from repeated observations depends 
on the type of star considered. We plot the scatter of repeated observations in 
Fig.~\ref{FIGrepeat_error_scatter_in_HR}, using the absolute $J$ magnitude as 
its label, in a manner similar 
to Fig.~\ref{FIGerror_scatter_in_HR} for the formal errors. 
The morphological properties of the scatter of repeated 
observations is similar to the formal errors, but the values 
involved are much smaller, except for stars with $M_J \simeq 1.0$,
where the scatters are similar.
The median error for Solar type dwarfs ($M_J \simeq 3.0$) is 
$\simlt 0.2$~mag. A similarly low scatter is reached also for 
red clump stars when using the Padova isochrones. 
The dashed lines in Fig.~\ref{FIGrepeat_error_scatter_in_HR} 
marking the number of repeated objects closely 
follow the distribution of the whole sample 
(Fig.~\ref{FIGerror_scatter_in_HR}). Again they peak in the 
regions of the smallest errors.

\subsection{Distance distribution}

\begin{figure}[hbtp]
\centering
\includegraphics[width=7.1cm,angle=270]{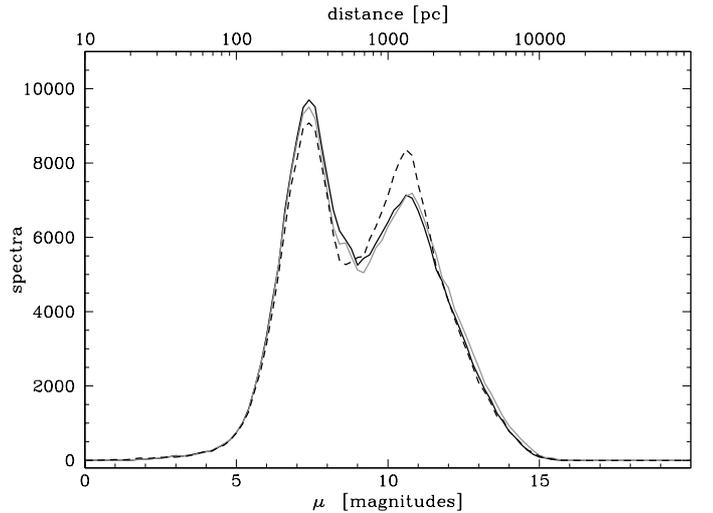}
\caption{Distribution of distance moduli ($\mu$) for the three 
isochrone sets: YY (black), Dartmouth (grey), and Padova
(dashed line). The bin size is 0.2 magnitudes. 
}
\label{FIGdistance_histogram}
\end{figure}

Fig.~\ref{FIGdistance_histogram} shows a distribution of distance 
moduli $\mu$. The results for YY and Dartmouth isochrones are very 
similar. The two peaks correspond to dwarfs at distances of $\sim 300$~pc 
and giants at $\sim 1.5$~kpc, with some stars being as far away as 10~kpc.
Calculations using Padova isochrones find more giants and their distribution
is more peaked. The difference is due to the red-clump evolutionary 
phase which is present only in this isochrone set.

\begin{figure}[hbtp]
\centering
\includegraphics[width=9.2cm,angle=0]{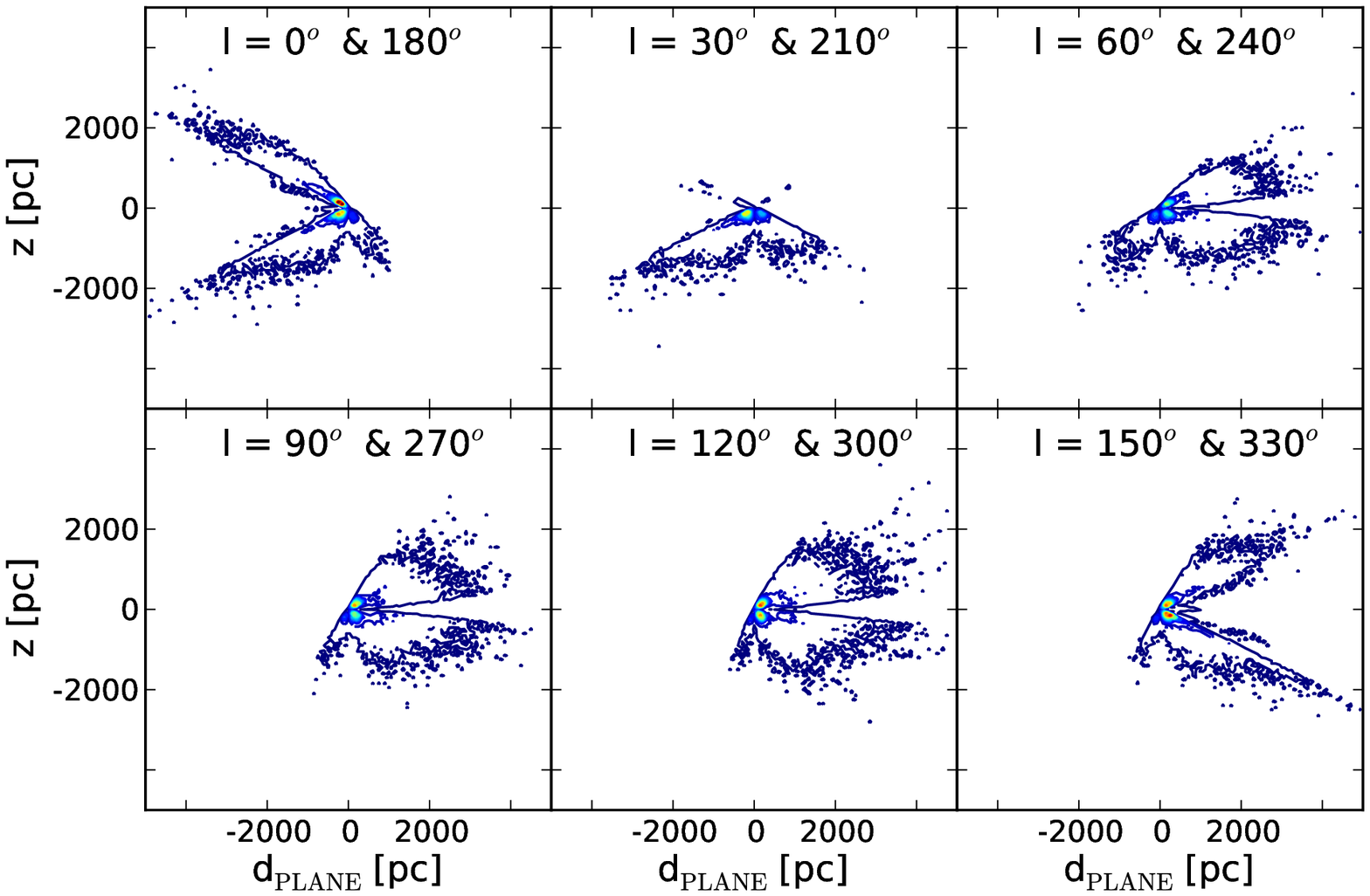}\\
\includegraphics[width=9.2cm,angle=0]{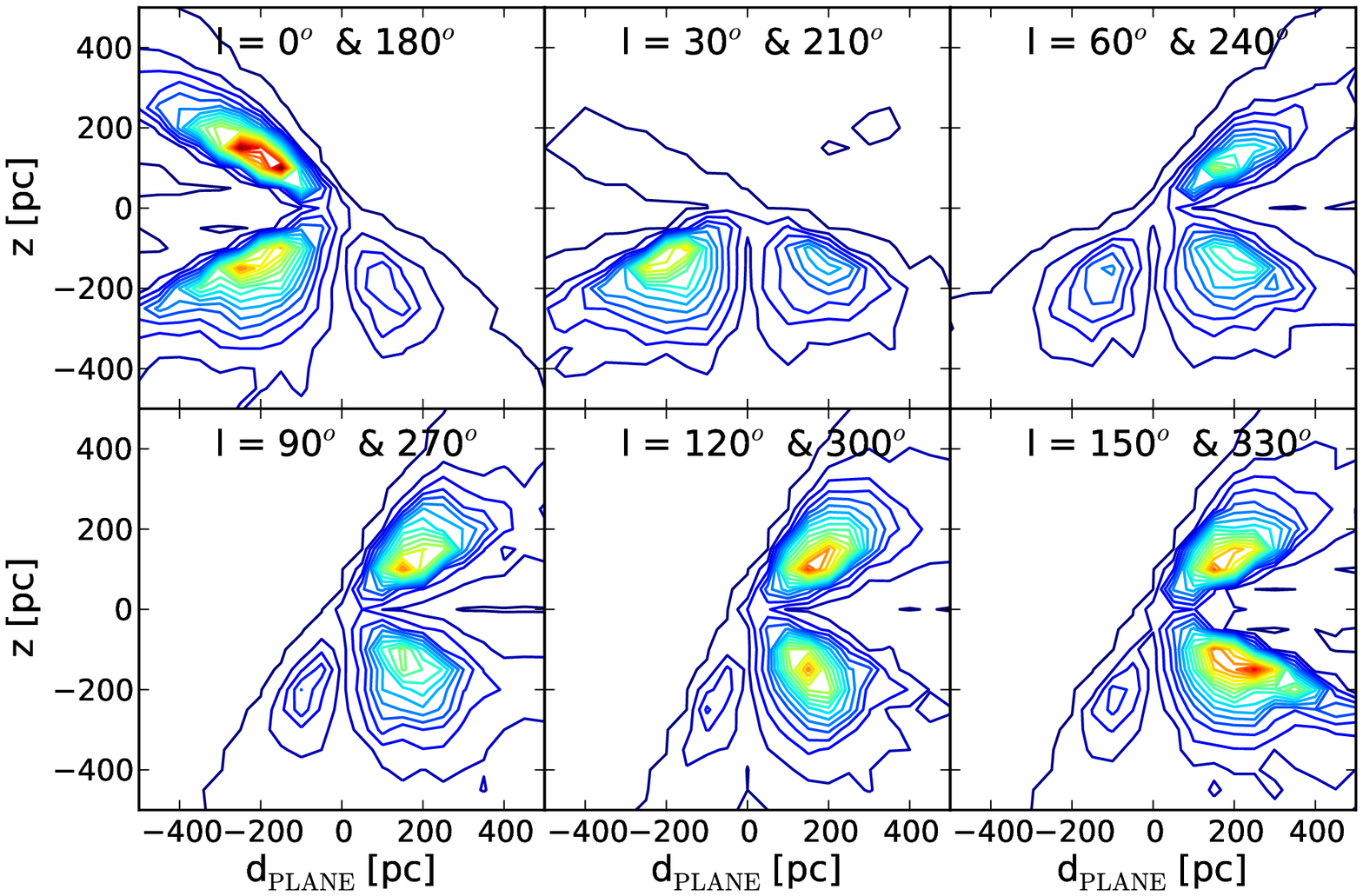}
\caption{
Galactic vertical cross-sections of the sample for distances 
derived using the Padova isochrone set. Each plot contains 
stars with Galactic longitudes within 15$^\mathrm{o}$ of the labelled 
values. $d_\mathrm{PLANE}$ is the cylindrical projected distance 
along the Galactic plane with negative values for stars within 
the Solar circle. $z$ is the distance from the Galactic plane. 
The density contours mark linearly spaced surface 
densities between 0.001 and 0.3 stars/pc$^2$. The upper panel shows 
the whole sample of $235\,064$ stars and the bottom panel 
shows an enlargement of the densest parts around the Galactic 
midplane.
}
\label{FIGtest115P}
\end{figure}

The observed sample, which is limited to the Southern Hemisphere,
is influenced by the Galactic structure. This causes a directional 
dependence of the derived distances. B10 plot the projections 
of the sample to the Galactic (x,y,z) coordinates (see Fig.~10 in B10). 
Here we make a more detailed plot (Fig.~\ref{FIGtest115P}), dividing 
the sample into $30^\mathrm{o}$-wide slices perpendicular to the 
Galactic plane. The upper set of graphs depicts
the general situation, and the bottom one is zoomed-in by a 
factor of 10. The Sun is located at the origin of the coordinate 
system, so its distance from the Galactic plane is neglected. 
About half of each graph is empty because we are 
limited to the southern skies. Also the general direction of the 
Southern galactic pole has been observed infrequently. This can 
be seen also in Fig.~21 of Z08. 

The zoomed-in set in Fig.~\ref{FIGtest115P} reveals 
an expected property. Many of the isodensity contours do 
not follow circular arcs around the Solar position. They tend 
to be horizontal, i.e.\ parallel to the Galactic plane. This is 
expected and is a direct consequence of us living in a disky 
galaxy. But the fact that we can see it in Fig.~\ref{FIGtest115P}
supports the notion that the measured distances are rather 
accurate. 

\begin{figure}[hbtp]
\centering
\includegraphics[width=6.9cm,angle=270]{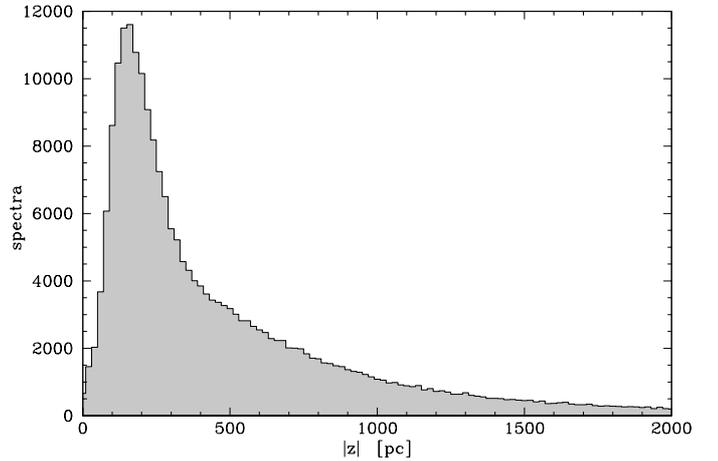}\\
\caption{
Histogram of distances from the Galactic plane if the Sun is 
assumed to be very close to it. The distances are based on the 
Padova isochrone set.
}
\label{FIGdisk_thickness}
\end{figure}

Motivated by this result we plot the histogram of distances from 
the Galactic plane (Fig.~\ref{FIGdisk_thickness}), again neglecting  
the Solar displacement off the plane. The result seems rather 
important. The RAVE sample generally avoids regions within 
$\sim 20^\mathrm{o}$ of the Galactic plane. So the histogram 
values increase until we hit the edge of the disk. The decreasing 
histogram values for stars with $|z| > 200$~pc are well represented 
by two exponential components, one of them with a very 
small scale-height. We believe such properties 
illustrate the capacity of RAVE to address the  
issues of Galactic structure, using its measurements of 
kinematics, temperature, gravity, metallicity, and now distance. 
But to draw secure conclusions about the properties of the thin 
and thick disks, even if immediately suggestive by the double exponent 
in Fig.~\ref{FIGdisk_thickness}, requires further study. In 
particular, a detailed assessment of possible selection effects 
and a comparison to models of the Galactic structure is needed. 
Therefore a complete discussion is deferred to a separate paper. 
%\citep[][ in preparation]{pasetto10}. 

\begin{table*}
%\begin{deluxetable}{ccllll}
%\rotate
%\tabletypesize{\small}
%\tablecaption
\caption{Description of the catalog of spectroscopic distances, available at the CDS.
}
\label{t:catalog}
%\tablewidth{0pt}
%\tablecolumns{5}
%\tablehead{
\centering
\begin{tabular}{ccllll}
\hline\hline
Column & Character & Format & Acronym    &Units               &Description  \\
number &           &        &           &                    &            \\
%}
%\startdata
\hline
%# OBJECTID        RA            DE           cluster clusterdist S2N  Breddist      YYdist   Dartdist Padovadist 
%#
%6.2f %7.3f %7.3f %7.3f %7.3f %7.3f %7.3f
1 &  1--17 & 17S  & Object-id&                    &Object name \\
2 & 19--30 & 12.8F&RA       &degrees             &Right Ascension (J2000.0)\\
3 & 32--43 & 12.8F&DE       &degrees             &Declination (J2000.0)\\
4 & 44--50 & 6.1F &RV       &km~s$^{-1}$         &Heliocentric radial velocity\\
5 & 52--57 & 6.1F &eRV      &km~s$^{-1}$         &Formal RV error\\
6 & 59--64 & 6.1F &pmRA     &mas~yr$^{-1}$       &Proper motion in RA$^1$\\
7 & 66--71 & 6.1F &epmRA    &mas~yr$^{-1}$       &Error in pmRA$^1$\\
8 & 73--78 & 6.1F &pmDE     &mas~yr$^{-1}$       &Proper motion in DE$^1$\\
9 & 80--85 & 6.1F &epmDE    &mas~yr$^{-1}$       &Error in pmDE$^1$\\
10& 87--92 & 6.3F &Jmag     & magnitude          &$J_{\rm2MASS}$ apparent magnitude\\
11& 94--99 & 6.3F &Kmag     & magnitude          &$K_{\rm2MASS}$ apparent magnitude\\
12&101--108& 8S   &Obsdate  &yyyymmdd            &Date of observation\\
13&110--119& 10S  &Fieldname&                    &Name of the field\\
14&121--123& 3G   &FibNum   &                    &Fiber number\\
15&125--130& 6G   &Teff     & K                  &Effective temperature\\
16&132--136& 5.2F &logg     & $\log($cm~s$^{-2})$&Surface gravity\\
17&138--142& 5.2F &$[$M$/$H$]$&                  &Metallicity\\
18&144--149& 6.2F &S$/$N    &                    &Signal to noise per pixel\\
19&151--157& 7.3F &$\mu$(YY)& magnitude          &Distance modulus (YY isochrones)\\
20&159--165& 7.3F &e$\mu$(YY)& magnitude         &Formal error in $\mu$(YY)\\
21&167--173& 7.3F &$\mu$(Dart) & magnitude       &Distance modulus (Dartmouth isochrones)\\
22&175--181& 7.3F &e$\mu$(Dart)& magnitude       &Formal error in $\mu$(Dart)\\
23&183--188& 7.3F &$\mu$(Padova) & magnitude     &Distance modulus (Padova isochrones)\\
24&190--196& 7.3F &e$\mu$(Padova)& magnitude     &Formal error in $\mu$(Padova)\\
\hline
%\enddata
\end{tabular}
\tablefoot{$^1$ If the item is not available its value is set to 9.99E8.}
%\end{deluxetable}
\end{table*}

\subsection{The Catalog}
\label{s:catalog}

The final catalog contains distances for $235\,064$ spectra 
of 210.872 different stars. These are RAVE spectra with spectroscopically 
determined values of radial velocity, surface temperature, gravity 
and metallicity. Furthermore these spectra have $\SN>20$ and were 
not identified as peculiar by the automated pipeline. 

A detailed description of the catalog is given in Table~\ref{t:catalog}.
Note that distances are given in distance moduli ($\mu$) and not 
as a plain distance ($d$). The reason is a nicer behaviour 
of errorbars. Errors of distance moduli are usually symmetric, 
while the corresponding distance errors are not. 
%The two values are connected via the usual relation
%\begin{equation}
%d = d_o 10^{0.2 \mu}
%\end{equation}
%where $d_o = 10$~pc.
%
The catalog entries will be made publicly available through the 
CDS service alongside the regular RAVE public data releases. Currently 
the latest is the second data release (Z08) with the third one close to 
publication \citep{siebert10}. Distances to RAVE stars are derived 
from values of stellar parameters and their adopted errors. Should 
these change in the future we will recalculate the distance set 
and report on the results by updating the corresponding astro-ph and
CDS entries.

\section{Discussion and conclusions}
\label{s:conclusions}

The paper uses spectroscopically determined values of stellar parameters
and the assumption that the star is a single object undergoing normal 
stellar evolution to determine its most likely distance. So the 
results are expected to be very useful for normal 
stars, while the values for peculiar objects (emission line stars, 
double line spectroscopic binaries) may suffer from systematic problems.
We note that peculiar objects present a minority in our sample, and 
we also try to detect them using an automated classification pipeline. 
The pipeline is conservative, still it classified $235\,064$ out of 
a total of $251\,364$ spectra with $\SN > 20$ as normal, i.e.\ a fraction 
of 93.5\%. Only the distances to such stars are published.

The present effort draws from the extensive experience gained during our 
initial work on the problem (B10). The method is similar but we introduce 
some refinements. First we note that isochrones calculated by different 
groups are not identical. So we give the distances using 
three recently published isochrone sets. The sets are those 
of Yonsei-Yale \citep{demarque2004}, Dartmouth \citep{dotter08}, 
and Padova \citep{bertelli08}. 
This leaves it to the discretion of the user 
to choose the isochrone set that is best suited to a particular 
problem. Next we make sure that isochrones are spaced linearly in 
time and locally weighted so that the prevalence of low-mass stars 
is taken into account (we showed that the last point is not crucial 
for the results we get). Also, we take into account that the errors 
of stellar parameter values vary across the parameter space. We 
confirm that the error values published in the latest RAVE data release
are rather conservative, still we prefer to be careful and avoid 
more optimistic error estimates as suggested by the repeated observations. 

The essential differences between our method and that of B10 is this. B10
jittered the data by the observational errors and sought the model star that
most closely matched each jittered data point, and then averaged the
individual distances obtained in this way. We calculate for 
every  model star the probability that it could give rise to the data, given
the known observational errors. Then we use these weights to average the
fluxes of the model stars and calculate the absolute magnitude of the star.
By using the closest match, the B10 scheme gives a large weight to 
rapid phases of stellar evolution which sweep the otherwise empty 
parts of the (temperature, gravity) plane. Here we assume 
that stars are observed at a random point of their lives, that 
less massive stars are more numerous, and that higher luminosity ones 
are surveyed in a larger volume of space.

The results have been tested using two sets of stars for which 
the distances are known a priori. These are Hipparcos stars with 
well determined values of the trigonometric parallax, and stars that 
are secure members of stellar clusters. Distances determined from 
RAVE spectra match the astrometric values to within $\simlt 22$\%, 
with a mean difference of less than 3\%. This is similar to the 
results already obtained by B10. Hipparcos stars with well determined
values of astrometric distance are almost exclusively dwarfs. So 
we need another set to check the distances derived for giant stars,
which represent about half of the RAVE sample. We used members of 
open and globular clusters observed by RAVE. We identified 
spectra of 12 stars that are certain members of 7 open clusters 
and one globular cluster. We note all isochrone sets yield distances 
which agree with cluster distances to within 10\% on average. 
But only the Padova isochrone set includes the red-clump phase, so
the distances derived using the Padova set are the preferred ones for 
giant stars. The results for individual stars show a scatter of 
$\sim 20$\% with respect to the published cluster distances. This is 
encouraging, as it shows that distances derived spectroscopically 
agree well with other methods up to a range of several kiloparsecs. 

The H-R diagram using the derived distances now appears rather 
clean, with two well defined groups of stars that present the 
majority of the RAVE sample (Fig.~\ref{FIGtest106}). These are 
Solar-like dwarfs with $\teff \sim 6000$~K and giants with 
$\teff \sim 4750$~K. About half of the latter are members of the 
red clump, with the rest being on the red-giant branch. We derive 
the absolute magnitude of the red clump as 
$M_J = -0.91 \pm 0.02$ and $M_K = -1.61 \pm 0.02$ which is 
consistent with the values from the literature. 

Distance errors have been estimated internally within the method 
and by using repeated observations of the same object. The internal 
estimates appear to be rather conservative and give median 
errors on distance of 33\%. 
The repeatability of derived distances is much better: the distances 
derived from individual spectra of the same object match at a level 
of 11\%.

\begin{figure}[hbtp]
\centering
\includegraphics[width=7.1cm,angle=270]{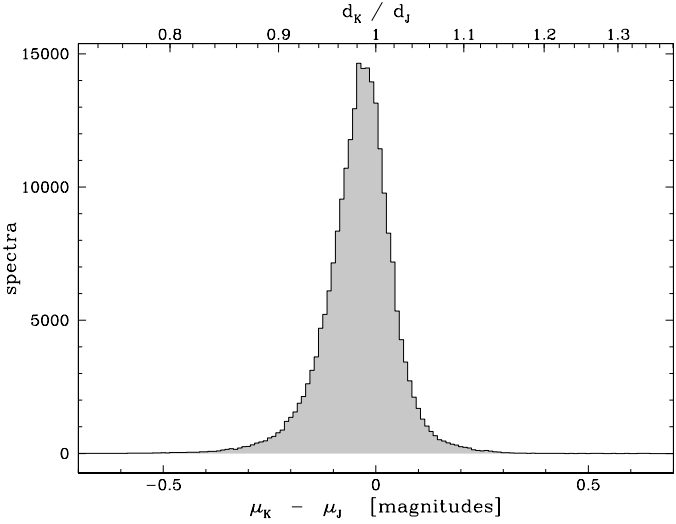}\\
\caption{
Comparison of distances (top axis) and distance moduli (bottom axis) for a 
calculation with Padova isochrones if using the $J_{\rm2MASS}$ or $K_{\rm2MASS}$ magnitudes. 
}
\label{FIGcomparedistJK}
\end{figure}

Derived distances are based on apparent magnitudes in the $J_{\rm 2MASS}$
band. We assume that interstellar reddening is negligible because the 
vast majority of RAVE stars lie at high Galactic 
latitudes ($|b| > 20^\circ$). Nevertheless it is interesting 
to compare the $J_{\rm 2MASS}$ results with the ones derived 
using the $K_{\rm 2MASS}$ band, which is even 
redder and so even less sensitive to reddening. So we calculated the 
distances to the whole sample using the $K_{\rm 2MASS}$ bandpass and the 
Padova isochrone set. Fig.~\ref{FIGcomparedistJK} compares the results
for both bandpasses. 
The average distance ratio is $\overline{d_K / d_J} = 0.983$ and the scatter is 
$\sigma(d_K / d_J) = 0.039$. Only 5\% of the spectra have 
$d_K / d_J < 0.92$ and 5\% of the spectra have 
$d_K / d_J > 1.04$. Comparison of our distances and the 
Hipparcos trigonometric parallax determinations 
(Sec.~\ref{s:Hipparcos})
gives similar results for both bands, with the scatter in the 
distance ratio being 0.25 for the $J$ band and 0.24 for the $K$ band. 
Also the results for stellar clusters are very similar. The 
scatter of spectroscopic distances around the cluster distances 
from the literature is 22\% for the $J$ band and 23\% for 
the $K$ band. 

One can test whether about 2\% smaller distances derived 
from the $K$ instead of the $J$ band are consistent 
with a decreased influence of interstellar reddening at 
longer wavelengths. From relations for extinction for 
the Solar energy distribution 
\citep[$A_J \simeq 0.887 E_{B-V}$ and $A_K \simeq 
0.382 E_{B-V}$,][]{munariadpspaper2}   
and from $\overline{A_K - A_J} \simeq 5 \log(0.983)$ we get 
$\overline{E_{B-V}} \simeq 0.07$. This is larger than 
the  values of integrated colour excess toward the South 
Galactic pole which range from $E_{B-V} = 0.018$ \citep{schlegel98} 
to $E_{B-V} = 0.04$ \citep{teerikorpi90}. The RAVE sample 
has $\overline{\csc |b|} = 3.22$, where $b$ is the Galactic lattitude.
So the difference between our average color excess and the values at 
the South Galactic pole can be explained with a simple slab model of the 
Galactic disk. 

We note that a comparison of distances derived using different passbands 
can improve our understanding of reddening at high 
Galactic lattitudes. This will be discussed in a separate paper, where 
we shall also address the rare cases of large differences in 
the derived distances which may point to spectral peculiarities. 
The photometric effect of interstellar reddening can then be confirmed 
spectroscopically from RAVE data by measurement of the equivalent width 
of the diffuse interstellar band at 8620~\AA\ \citep{munari_DIBS}.  
We conclude that distances derived using either the $J$ or the $K$ 
band are very similar, but a bit smaller measurement errors of 
apparent magnitudes favour the $J$ band as our baseline.

The main result of this paper is a catalog of newly derived stellar 
distances. The paper discusses distances to $235\,064$ stars. The 
list will increase with time as the RAVE survey progresses. The 
public version of the catalog will be accessible through the 
CDS service, with its size growing along with the public RAVE 
data releases. Note that this is not the only current effort to 
derive distances to RAVE stars. Our results are based on the 
work of B10, and there is also a separate project of distance 
derivation using a Bayesian approach \citep{burnett10}. We believe 
the availability  of 
complementary computations will be of assistance to users, who can then
select those best suited 
to their particular problem. 

The present results place RAVE dwarfs at typical distances of 
$\sim 300$~pc, and giants at 1 or 2~kpc, even though there are 
objects 10~kpc away. In fact RAVE reached even the brightest LMC 
stars \citep{munari_LBV}. So the RAVE survey is well placed between 
the more local Geneva Copenhagen  Survey 
\citep[GCS,][]{nordstrom2004,holmberg09} and the stellar component 
(SEGUE) of the Sloan Digital Sky Survey \citep{abazajian09}. 
Two thirds of the GCS stars are closer than 100~pc and 95.5\% 
are closer than 200~pc from the Sun, while most of the SEGUE stars are 
members of the halo. For example
the full sample of SDSS/SEGUE DR-7 calibration stars contains
only $\sim 2400$ stars within 1~kpc from the Galactic 
plane \citep{beers09}. These results make the RAVE survey ideal 
to study the properties of the Galactic thin and thick disks. 
The flatness of the stellar distribution can be seen also directly 
from the shape of the isodensity contours in cross-sections 
perpendicular to the galactic plane. This confirms the capability 
of the RAVE survey to address some of the most fundamental questions 
of Galactic astrophysics. The possibility to upgrade the RAVE 
metallicity measurements with individual element abundances 
\citep{boeche10} only reinforces such a conclusion. 
Individual applications of this dataset with full 6-dimensional 
kinematical and physical information will be discussed in separate 
papers.

\begin{acknowledgements}
Funding for RAVE has been provided by: the Anglo--Australian 
Observatory; the Astrophysical Institute Potsdam; 
the Australian National University; the Australian Research Council; 
the French National Research Agency; the German Research foundation; 
the Istituto Nazionale di Astrofisica at Padova; 
The Johns Hopkins University; 
the National Science Foundation of the USA (AST-0908326); 
the W.M. Keck foundation; the Macquarie University; 
the Netherlands Research School for Astronomy; 
the Natural Sciences and Engineering Research Council of Canada; 
the Slovenian Research Agency; the Swiss National Science Foundation; 
the Science \&\ Technology Facilities Council of the UK; 
Opticon; Strasbourg Observatory; and the Universities of Groningen, 
Heidelberg and Sydney.
The RAVE web site is at http://www.rave-survey.org.
\end{acknowledgements}

\bigskip
\appendix{\bf Note added in proofs}
\bigskip

New values of stellar parameters became available after acceptance of the paper.
The original results are based on the parameter determination pipeline as described in
\citet{Z08}, while the derivation of new values is explained in \citet{siebert10}. Another recent 
progress is a redetermination of errors of stellar parameters across the H-R diagram.
Instead of the conservative external errors reported in \citet{Z08} we could use an extensive 
RAVE subset where a certain star was observed more than once. Errors on stellar 
parameters were estimated from the scatter of independent parameter 
determinations for a given star and averaged over stars in the given bin in temperature 
and gravity \citep{siebert10}. This internal error seems trustworthy, as there 
are no appreciable zero-point offsets between our stellar parameter values and 
those of the reference datasets \citep[see Table~5 in ][]{Z08}. The distance moduli 
have been calculated for the Padova isochrone set using the new values of stellar 
parameters. The assumed errors were either the conservative values used in the accepted 
version of the paper or the ones determined from repeated observations. 
Both distance moduli and their errors add four columns to the catalog of spectroscopic 
distances, available via CDS. 

New values of stellar parameters yield distances which are generally closer to the reference 
values than the ones reported in the main paper. The scatter ($\sigma$) of the 
spectroscopic ($d$) to trigonometric ($d_\varpi$) distance ratio is 0.28 and 0.22 if 
the external and internal error estimates are used, respectively. This test was 
done on the same sample of Hipparcos stars as the one reported in Table~1 where the 
corresponding scatter equals 0.25. A similar test can be performed using members 
of star clusters. The scatter of the spectroscopic to cluster distance ratio is 0.29 
for the distance calculations using Padova isochrones and reported in the main paper.  
The new values of stellar parameters diminish the scatter to 0.205. This value is 
obtained for both choices of error estimation: the conservative external error 
estimates and the internal ones obtained from repeated observations. We conclude that the best match between 
our spectroscopic distances and the reference values is obtained when using the 
Padova isochrone set, new values of stellar parameters and errors determined from 
repeated observations.

\end{document}